\documentstyle[prb,aps,epsf]{revtex}
\tighten
\begin{document}
\draft 
 
\title{Quantum Chaos in Quantum Wells }
\author{E.~E.~Narimanov   and  A.~Douglas ~Stone}   
\address{Applied  Physics,  Yale  University,  P.O.  Box  208284,  New 
Haven CT06520-8284}
\date{\today}  
\maketitle

\section{Introduction}

Although the field of quantum chaos found its first experimental
challenges in atomic physics, during the past eight years several
systems in condensed matter physics have been identified and studied
from this point of view.  In all of these cases the low-temperature
electronic conduction properties of semiconducting heterostructures
have been measured and analyzed to infer properties of the electronic
spectrum and wavefunctions which may reveal quantum manifestations of
chaos, or of the transition to chaos.  Of these experiments, those
which have involved patterning of two-dimensional electron gas systems
to create more complex geometric scattering have in general not been
amenable to detailed semiclassical analysis because the precise nature
of the underlying confinement potential is not known.  In contrast, the
series of experiments performed at Nottingham University \cite{Nott}
and Bell Laboratories \cite{Bell} which we will review and analyze
here, involve tunneling through a quantum well in high magnetic field,
for which the classical dynamics within the well can be accurately
modelled and related to experimental observables.  Moreover this system
is unique in that three {\it in situ} experimental parameters, magnetic
field, B, bias voltage V, and the angle between the electric and
magnetic field, $\theta$ can be continuously varied so as to generate a
transition from integrable to (essentially) fully chaotic dynamics.  As
such the system provides an ideal testing ground for semiclassical
methods; we will find that is also raises some new questions in
non-linear dynamics of some general theoretical interest.  We will
review and summarize below our theoretical study of this system
\cite{tiltprb,tiltscar,tiltsc},  while also extending our earlier semiclassical
treatment of the problem \cite{tiltsc}.

We begin with some essential background.  It is possible to fabricate
atomically-flat layered materials, typically of the semiconductor GaAs
and the insulator AlGaAs, such that the band-gap varies in a controlled
manner along the longitudinal (growth) direction.  This creates an
effective one-dimensional potential for the electrons along this
direction which confines the free electrons or holes (which are
introduced by doping) in the semiconducting layers, known as ``quantum
wells''.  In the experiments in question, a thin ``emitter'' layer is
fabricated with only one longitudinal quantum level filled by doping,
followed by a barrier, a wide ($\sim 120$ nm ) empty quantum well, a
second barrier, and a conducting ``collector'' region.  Thus one has in
the longitudinal direction the potential profile of a ``resonant
tunneling diode''.  As the voltage across the wide well is varied,
states in the well can become aligned with the emitter level,  and
resonances in the tunneling current through the double-barrier
structure will be measured if emitter and well states are coupled.  In
the experiments we will discuss, a large magnetic field is imposed,
initially perpendicular to the layers, and has the effect of quantizing
the electronic states within the layers into highly degenerate Landau
levels.  The electronic density in the emitter is such that typically
only the lowest (n=0) Landau level is full, and so an exact selection
rule forbids tunneling through any but the n=0 states of the well.  
Since both the emitter and well states have the same linear energy
variation with B, one measures a series of peaks in the I-V
characteristic of the device with voltage spacing corresponding to the
longitudinal quantum states of the well, independent of the value of the
magnetic field. Resonant tunneling diodes (RTDs) of this type have been
well-studied for more than a decade.

The novel features of the more recent experiments arose when the
magnetic field was tilted with respect to the electric field
(tunneling) direction.  We choose coordinates such that the normal
direction to the layers is the $z$-direction, and the tilt angle is in
the y-z plane, so that ${\bf B} = \cos \theta \hat{z} + \sin \theta
\hat{y}$ (see Fig. \ref{fig:schematic}).  
It was observed by Fromhold et al. \cite{Nott}
that as the tilt angle was varied at high magnetic field (B=10T), the
frequency of resonance peaks in the I-V would abruptly change at
certain values of the voltage (typically this frequency approximately
doubled). These workers  initially attributed this frequency-doubling
to a density of states (DOS) effect within the well.  They showed
numerically that at large tilt angles the electron dynamics in the well
is almost completely chaotic, and that there exist new unstable
periodic orbits which collide with both the emitter and collector
barriers. These orbits were supposed to give rise (at certain voltages) to 
additional DOS
oscillations via Gutzwiller's Trace Formula.  A later numerical
analysis of the quantum states within the well found that, unlike the
usual level-clustering expected from the Trace Formula, in this case
individual levels were scarred by the relevant periodic orbits once
within each period of the Gutzwiller oscillation.  Moreover, upon
solving for the tunneling current by a quantum transfer matrix method
it was found that in many cases these scarred levels carried
essentially all of the resonant tunneling 
current\cite{Fromhold_scars_prl,Fromhold_scars_nature}.

Additional interesting features of this system were found when Muller
et al. \cite{Nott} at Bell Labs undertook a careful study of the
entire parameter space of magnetic field ($0 < B < 12$T), voltage 
( $0 < U < 1$V) and tilt angle, focusing particularly on the transition
region to chaos.  They analyzed this large data set by plotting the
peak locations in the B-V plane for fixed  tilt angles between eleven
and forty-five degrees.  This representation of the data revealed a
regions of peak-doubling and tripling  which evolved in a complex
manner with increasing tilt angle.  These experiments motivated
Shepelyansky and Stone \cite{ss} to analyze the global phase-space
structure of electron motion within the well.  Using a simplified model
which neglected collisions with the emitter barrier they were able to
estimate via the Chirikov resonance overlap criterion \cite{Chirikov}
the transition to global chaos in this system as a function of
B,V,$\theta$.  They also suggested that the onset of peak-doubling in
the system was due to bifurcations of the simplest ``traversing
orbit'', an interpretation adopted and amplified upon by Muller et al.
\cite{Bell}.   However the classical analysis of Shepelyansky and
Stone was not suitable for a developing a quantitative semi-classical
theory, since it did not describe orbits which collide with both
barriers, which dominate the tunneling spectrum according to both
semiclassical theory \cite{tiltsc} and exact quantum analyses
\cite{Fromhold_scars_prl,Fromhold_scars_nature,Monteiro,Saraga}.

The initial experimental and theoretical/numerical results on this
system raised a number of very interesting questions:

1) Is it possible to formulate a semiclassical theory of the tunneling
current which involves only real (as opposed to complex) orbits within
the well?  If so, would such a theory only involve periodic orbits (and
not all closed orbits as one finds in diamagnetic hydrogen), and would
all or only a subset of the periodic orbits be important?

2) The classical dynamics of electron motion within the well was
observed numerically to have some unusual features.  Certain regions of
phase space became highly chaotic well before a KAM analysis would
predict.  Moreover, periodic orbits which connected the emitter and
collector barriers were observed to disappear by backwards bifurcations,
but it was rarely if ever possible to find the bifurcations in which
these orbits were born.  What was causing these unusual properties?

3) Was the scarring of chaotic wavefunctions unusually common in this
system?  If so, what was the origin of this strong scarring, and why
were only certain unstable periodic orbits reflected in the observed
scarred wavefunctions?

We attempt to give answers to all of these questions below, based on 
analysis of
the classical dynamics and semiclassical methods.  We note that this 
system has
been extensively studied numerically, both classically via the 
Surface-of-Section method
\cite{ss,tiltprb,Fromhold_prbrc}, and 
quantum-mechanically by analysis of scaled 
spectra \cite{Monteiro,Saraga} and tunneling current  
\cite{Fromhold_scars_prl,Fromhold_scars_nature,Monteiro,Saraga}.  
In addition to our own semiclassical treatment of the problem \cite{tiltsc}, recently
there has been an independent work by Bogomolny and Rouben which presents
two possible semiclassical tunneling formulas for this system \cite{BR}.  The 
first of these
expressions neglects the structure of the emitter wavefunction and leads to a result 
which differs substantially from our periodic orbit formula, 
Eq. (\ref{w_osc}) below.  In Ref. \cite{Monteiro1} the second 
was shown below to be equivalent to our Eq. (\ref{w_osc}), first presented in
Ref. \cite{tiltsc}, when similar assumptions were made
about the emitter state (we present a similar proof and detailed comments below).  

Recent comparisons of this
formula with exact quantum calculations \cite{Monteiro1} and with experimental
data \cite{tiltsc,Monteiro1} find good agreement.
However, the periodic orbit formula was shown to lose accuracy and in some cases
fail badly in narrow 
regions\cite{narrow} near bifurcations, when the relevant periodic orbit
does not yet exist or is not isolated.  This failure is a manifestation of the ``ghost'' effect \cite{ghost_ref}.  The relevance of this effect to the tilted well was first pointed out
by Monteiro et al. \cite{Monteiro} who have recently shown
that an adequate semiclassical description in this regime can be achieved by
using complex dynamics \cite{Monteiro2}.
Alternatively, we show below that one can use a  
representation of the tunneling current in terms of closed orbits 
(Eq. 14, section IIB), which involves only real orbits within the well and provides a
remarkably precise quantitative description of the ghost effect (see section IVB).
It is important to note that for the tilt angles of interest the spectrum cannot be 
calculated perturbatively from 
the $\theta = 0$ limit, so that the {\it only } analytic option is a semiclassical 
approach, exploiting the fact that we are interested in states far above 
the ground state of the well, and the large level-broadening as discussed 
below.

\section{ Semiclassical Theory of Resonant Tunneling}

\subsection{Bardeen Approximation}

In standard tunneling theory the current will be proportional to the
transmission coefficient from the emitter to the collector\cite{Payne}.  
This
presents an immediate complication for a semiclassical formulation of
the problem because the transmission coefficient involves the Green
function across classically forbidden regions and hence would involve
complex paths traversing the barriers.  
However another aspect of the physics suggests an
alternative and simpler approach.  The electrons are injected by
tunneling from the emitter state into the well under high bias and far
from equilibrium.   Hence they will emit optic phonons and lose energy
over a time scale of 0.1ps corresponding to 4-5 collisions with the
barriers, a time much shorter than the total tunneling time through the
well which (from the measured tunneling current) corresponds to $\sim 
100-1000$ collisions.  Therefore there is a substantial level-broadening, 
corresponding to several
level-spacings within the well, and the tunneling is sequential, not
coherent.  Under such circumstances, in which correlations between
tunneling into and out of the well can be ignored, one may employ the
Bardeen tunneling formalism\cite{Bardeen}. 

In Bardeen's ``tunneling hamiltonian'' approach, the system is treated
as a sum of three isolated parts -
the emitter, the quantum well, and the collector. Tunneling is considered 
as a perturbation, which introduces a nonzero coupling between these 
subsystems. The tunneling rate is then calculated using the Fermi Golden 
Rule, with the coupling matrix element expressed in terms of the 
wavefunctions of the isolated subsystems. One can then write a system
of rate equations, expressing the balance of tunneling currents to and from
the quantum well, and determine the nonlinear conductance of the system. 

If the transmission coefficients of the emitter and collector
barriers are of the same order of magnitude, a substantial space charge
is accumulated in the quantum well. As a result, the 
electron-electron interactions in the well become relevant, and it is not
possible to connect the properties of resonant
tunneling spectra to the classical mechanics of a {\it single} electron
in the well. In order to probe the single-electron dynamics in the 
quantum well, the charge build-up in the well must be prevented. That 
can be achieved only when the width of the collector barrier is
much thinner than the width of the emitter barrier, so that the 
tunneling rate from the well to the collector is much larger than the 
tunneling rate from the emitter to the well.  In such a case we may set 
the 
occupation number of states in the well to zero in the rate equations for 
the
current.

In this regime the rate limiting step is the
tunneling from the emitter to the well, and the tunneling current is 
therefore 
proportional to the corresponding  {\it tunneling rate} 
(probability of tunneling per unit time). If there is only one
energy level occupied in the emitter 2DEG (which is the case 
in the experiment at high magnetic fields $B \geq 5$T), the 
tunneling current $j$ is given by
\begin{eqnarray}
j = n_e e W^{e \rightarrow w} \label{jw}
\end{eqnarray}
where $n_e$ is the surface concentration in the emitter layer.
The Eq. (\ref{jw}) can be easily generalized to the case,
when the emitter layer contains several occupied single-electron
levels.

The tunneling rate from the emitter to the well
\begin{eqnarray}
W^{e \rightarrow w} = \frac{2 \pi}{\hbar} 
\sum \left| M^{e \rightarrow w} \right|^2 \delta\left(\varepsilon_w - 
\varepsilon_e\right)
\label{golden_rule}
\end{eqnarray}
where the summation is performed over all energy levels in the well,
and the coupling matrix element
\cite{Bardeen,Payne} is given by
\begin{eqnarray}
M^{e \rightarrow w} = \frac{\hbar^2}{2{m^*}} \int_{S} 
\left( \Psi^{w} {\bf \bigtriangledown} \Psi^{e} -
\Psi^{e} {\bf \bigtriangledown} \Psi^{w} \right) d{\bf S}
\label{Bardeen_me}
\end{eqnarray}
with the wavefunctions $\Psi_e$ and $\Psi_w$ corresponding respectively
to {\it isolated} emitter and {\it isolated} well.

The integration in (\ref{Bardeen_me}) can be performed over any surface
inside the barrier, including the boundary of the barrier itself. We 
choose 
to perform the integration over the ``inner'' boundary, which is 
classically 
accessible to the electrons trapped inside the well. The latter allows the 
semiclassical treatment of the wavefunctions of the isolated well using the
Gutzwiller path integral approach \cite{Gutzwiller}.

The  expression (\ref{Bardeen_me}) involves both the wavefunction of the
isolated well $\Psi^w$ and it's normal derivative 
$\bigtriangledown \Psi^w$, taken at the 
surface of the barrier ($z = 0$) . In the limit of large height of the
emitter barrier $U_0$ corresponding to the experimental setup
\cite{experimental_values} the
normal derivative of the wavefunction does not depend on $U_0$, while 
for the wavefunction itself calculated at the plane of the barrier the
dependence on $U_0$ is crucial : $\Psi^w\left(z = 0\right) \sim
\exp\left( - \sqrt{U_0/\varepsilon_i} \right)$, where the ``injection
energy'' $\varepsilon_i$ is defined as the difference between the total
energy $\varepsilon$ and the potential drop across the well $eV$.
Since both terms in Eq. (\ref{Bardeen_me}) are of the same 
order of magnitude, Eq. (\ref{Bardeen_me})
requires an adequate representation of both the well wavefunction and it's
normal derivative. 

The standard
semiclassical path-integral approach, which includes only real classical
trajectories, does not distinguish between infinite and 
finite potential barriers, 
and cannot  be used for the calculation of the 
exponentially small, but non-zero value of the wavefunction at the plane
of the barrier. This makes it complicated to apply directly
the Eq. (\ref{Bardeen_me}) 
to the semiclassical description of the resonant 
tunneling. Instead, we first reduce the expression (\ref{Bardeen_me})
to the form, which involves only the normal derivatives of the wavefunction
of the isolated quantum well \cite{explain_where_U0_is}. In  the leading 
order
in $\hbar \omega_c/U_0$, $\varepsilon_i/U_0$ and $eE a_e/U_0$ 
(where $a_e$ is the width of the emitter barrier), the 
matrix element (\ref{Bardeen_me}) can be expressed as
\begin{eqnarray}
M^{e \rightarrow w}_{nk} = \frac{\hbar^2}{{m^*}} 
\int_{-\infty}^{\infty} \ dx 
\int_{-\infty}^{\infty} \ dy 
\ \Psi^{e}_n(x,y,0) \left. 
\frac{\partial\Psi^{w}_k(x,y,z)^*}{\partial z}\right|_{z=0}
\label{me_approx}
\end{eqnarray}
To obtain Eq. (\ref{me_approx}) we neglect the effects of the electric field
and the transverse component of the magnetic field in comparison
with $U_0$ 
in barrier region $-a_e < z < 0$.
Note, that for the untilted system ($\theta = 0$)  the approximation 
(\ref{me_approx}) gives {\it exactly} the same result as the standard 
WKB method.

With the choice of the vector potential 
${\bf A} = (- B y \cos(\theta) + B z \sin(\theta)) {\bf {\hat x}}$ 
the system possesses translational symmetry in the $x$-direction.
Therefore, the dynamics within the well can be reduced to two degrees of 
freedom (see below)
$y,z$, \cite{Bell,ss,Monteiro,tiltprb} with an effective potential
$V\left(y,z\right)$. The well wavefunctions in Eq. (\ref{me_approx}) 
can be
re-expressed in terms of the Green function of the isolated well,
$G(y_1,z_1=0;y_2,z_2=0;\varepsilon)$:
\begin{eqnarray}
W^{e \to w} & = & - \frac{2 \hbar^3}{{m^*}^2} \Im \int\int \ dy_1 \ dy_2 \ 
\Psi_n^e\left(y_1, 0\right)  
\Psi_n^e\left(y_2, 0\right)^*  
\left.
\frac{ \partial^2 G\left(y_1, z_1; y_2, z_2; \varepsilon_n\right) }
     { \partial z_1 \partial z_2 }
\right|_{z_1 = z_2 = 0}
\label{w_green}
\end{eqnarray}

The emitter state $\Psi_e$ in Eq. (\ref{me_approx}) involves only the 
few lowest single-particle levels and can be
calculated accurately using a variational approach \cite{Ando}. The 
``injection function'' $\Psi_e\left(y,0\right)$ is then a linear 
combination of the lowest few Landau level wavefunctions $\Phi_n$, and
is peaked at an injection point $y_i \approx
( a_e + 1.13 (\hbar^2 / {m^*}eE)^{1/3} ) \tan\theta$, and has spatial
extent of order the magnetic length,
$l_B \equiv \sqrt{\hbar/eB\cos\theta} \sim \hbar^{1/2}$:
\begin{eqnarray}
\Psi_e\left(y,0\right) & = & l_B^{-1/2} \sum_n c_n \Phi_n\left( 
\frac{y - y_i}{l_B} \right) 
\label{emitter_state}
\end{eqnarray}

\subsection{Semiclassical Approximation: The Closed Orbit Formula}

In the semiclassical approximation \cite{Gutzwiller}, the derivative 
of Green function $\partial^2_{z_1 z_2}
G(y_1,z_1=0;y_2,z_2=0;\varepsilon)$ is
determined by {\it all} classical trajectories connecting the points
$(y_1,0);(y_2,0) \equiv (y - \Delta y / 2, 0); (y + \Delta y / 2, 0)$,
and, as shown in Appendix A, can be expressed as:
\begin{eqnarray}
\left.
\frac{ \partial^2 G\left(y_1, z_1; y_2, z_2; \varepsilon_n\right) }
     { \partial z_1 \partial z_2 }
\right|_{z_1 = z_2 = 0}
& = & 
\frac
{8 \pi}
{\left( 2 \pi i \right)^{3/2} \hbar^{7/2}  }
\sum_\gamma \ 
\left(p^\gamma_i\right)_z \left(p^\gamma_f\right)_z
\ D_\gamma^{1/2} \ 
\exp\left[ \ 
i \
\frac{
S_\gamma\left(y - \Delta y/2, 0; y + \Delta y/2,0;
\varepsilon\right)
}
{\hbar} \ 
\right]
\label{green}
\end{eqnarray}
where $S_\gamma\left(y - \Delta y/2, 0; y + \Delta y/2,0;
\varepsilon\right) \equiv S_{\gamma}$ is the action of the classical 
trajectory\cite{Weyl} indexed
by $\gamma$, 
$D_\gamma$ is the appropriate amplitude \cite{Gutzwiller},
and
$p^\gamma_i$ and $p^\gamma_f$ correspond to
the initial and final momenta of the trajectory.
Substituting (\ref{green}) into (\ref{w_green}), and introducing
the effective broadening $\hbar/\tau_{\rm opt}$ of the energy levels
in the well, for the oscillatory part of $W_{e \to w}$ we obtain:
\begin{eqnarray}
W_{\rm osc} & = &
\int dp_y \int dy \ f_W\left(y,p_y\right)
\ \sum_\gamma \
\frac{\left(p^\gamma_i\right)_z \left(p^\gamma_f\right)_z}
{ \left({m^*}\right)^2 }
\int d \Delta y \ 
\Re   \left\{
\frac
{8 D_\gamma^{1/2}}
{\sqrt{2 \pi \hbar i}  }
\exp\left[
- \frac{t_\gamma}{\tau_{\rm opt}}
+
i
\frac{
S_\gamma
- p_y \Delta y}
{\hbar}
\right]
\right\}
\label{unclosed}
\end{eqnarray}
where $ t_\gamma$ is the classical propagation time of the
trajectory $\gamma$, and the factor
$\exp\left( - {t_\gamma}/{\tau_{\rm opt}} \right)$
represents the effect of phonon emission. This factor suppresses the
coherent contributions of trajectories longer than $\tau_{\rm opt}$.

We have also introduced in Eq. (\ref{unclosed})
the Wigner transform $f_W^e$ of the emitter wavefunction
 \begin{eqnarray}
f_W^e\left(y, p_y\right) \equiv h^{-1} \int d \Delta y
\ \Psi_e\left(y - \Delta y/2, 0 \right)
\ \Psi_e^*\left(y + \Delta y/2, 0 \right)
\ \exp\left( i p_y \Delta y/ \hbar \right)
\label{define_wigner}
\end{eqnarray}
which describes the distribution in transverse 
position and momentum of electrons
injected into the well.  Since $\Psi_e (y)$ has a width $\sim l_B$,
the integrand in Eq. (\ref{unclosed}) will be small for 
$\Delta y > l_B \sim \hbar^{1/2}$.

Assuming that $M$ Landau levels contribute to the emitter state
wavefunction, then the Wigner function
\begin{eqnarray}
f_W^e\left(y, p_y\right) & = & \frac{1}{h l_B} 
\sum_{k',k'' = 0}^{M-1} {
\cal U}_{k' k''} 
\left( \frac{y - y_i}{l_B} \right)^{k'} 
\left( \frac{p_y l_B}{\hbar} \right)^{k''} 
\exp\left\{ - \left(\frac{y - y_i}{l_B}\right)^2 - 
             \left(\frac{p_y l_B}{\hbar}\right)^2 \right\}
\label{emitter_wigner}
\end{eqnarray}
where the coefficients ${\cal U}_{k' k''}$ can be obtained from the
relation
\begin{eqnarray}
\sum_{k',k'' = 0}^{M-1} {\cal U}_{k' k''} 
\left( \frac{y - y_i}{l_B} \right)^{k'} 
\left( \frac{p_y l_B}{\hbar} \right)^{k''} 
& \equiv & 
\sum_{n',n'' = 0}^{M-1} \ c_{n'} c^*_{n''} I_{n' n''}
\label{define_u}
\end{eqnarray}
where
\begin{eqnarray}
I_{n' n''} & = & 
(-1)^{ {\rm min}\left(n',n''\right)} \ 
2^{1 + \frac{\left|n' - n''\right|}{2} } \  
\frac{ {\rm min}\left(n',n''\right)! }
     { \sqrt{ n'! \ n''!} } \ 
\left( \frac{y - y_i}{l_B} - i \ {\rm Sign}\left(n' - n''\right)
\frac{p_y l_B}{\hbar}
\right)^{\left|n' - n''\right|}
\nonumber \\
& \times & 
L_{ {\rm min}\left(n',n''\right) }^{\left| n' - n'' \right|}
\left( 2 \left(\frac{y - y_i}{l_B}\right)^2 + 
       2 \left(\frac{p_y l_B}{\hbar}\right)^2 \right)
\label{integral} 
\end{eqnarray}
and $L_n^m$ is the generalized Laguerre polynomial.

Consider first the integration over $\Delta y$ in Eq. (\ref{unclosed}).  
Since
$\Delta y \sim \hbar^{1/2}$, in the
semiclassical limit $\hbar \to 0$, one can expand the scaled 
action $S_{\gamma}
(y-\Delta y/2,0;y + \Delta y/2,0;\varepsilon)/\hbar$ in $\Delta y$, 
retaining only terms up to second order, and perform
exactly the resulting Gaussian integral.  Alternatively, we can employ the
approach
of Berry \cite{Berry} and perform this integration by stationary phase, 
which
will initially lead to an expression in terms of non-closed orbits which
satisfy
the ``mid-point rule'', and then re-express this answer to the same 
accuracy
in $\hbar$ using $\Delta y \sim \sqrt{\hbar}$ to arrive at the same result
involving
only families of closed orbits:
\begin{eqnarray}
W_{\rm osc} & = & 
\frac{16}{{m^*}} \int dy \int dp_y f_W\left(y, p_y\right)
\sum_\alpha
\sqrt{ 
\frac{(p^\alpha_i)_z (p^\alpha_f)_z}
     { m_{11}^\alpha + m_{22}^\alpha + 2 }
}
\exp\left[ - \frac{i}{\hbar} 
\frac{2 m_{12}^\alpha}{m_{11}^\alpha + m_{22}^\alpha + 2} 
\left(p_y - \bar{p}^\alpha_y\right)^2 \right]
\exp\left[i \frac{S_\alpha\left(y,0;y,0;\varepsilon\right)}{\hbar} \right]
\label{closed}
\end{eqnarray}
where $M_\alpha \equiv (m^{\alpha}_{ij})$ is the
$2\times2$ monodromy matrix\cite{Gutzwiller}, defined via the 
linearization
of the Poincar\'e map near the closed orbit $\alpha$ , and 
calculated at the contact point at the emitter barrier.  

The $y$-integration now involves the rapidly-varying phase
$\exp [i S_{\alpha}(y,y)/ \hbar]$ for closed orbits beginning and ending
at $y$. This phase is stationary for the periodic orbits, and we therefore
expect the periodic orbits to determine the tunneling current.
However, $f_W(y)$ varies on the
same spatial scale $\sim \hbar^{1/2}$, and we therefore cannot immediately
perform the $y$ integral by stationary phase (as is done to
derive the trace formula for the total density of states\cite{Gutzwiller}).
The failure of such an approach in this case reflects a more general principle of
semiclassical theory; experimental observables typically do not depend only on
the density of states, but also upon coupling matrix elements.  The classical orbits
which will contribute then depend on the nature and spatial extent of the coupling.
In fact in the well-known problem of 
diamagnetic hydrogen, the ``source function'' 
analogous to $f_W (y)$
varies on a scale much smaller than the relevant phase, being
effectively a delta-function, and the resulting semiclassical theory
involves {\it all} closed orbits at the source point \cite{Delos} and not periodic 
orbits.  The current situation is intermediate between the absorption spectrum for diamagnetic hydrogen and a pure density of states measurement; a periodic orbit
sum can describe the tunneling spectrum quantitatively, but a closed orbit formula
(Eq. (\ref{done_y}) below) is more accurate near bifurcations.  
However the periodic orbit
formula cannot be obtained by stationary phase integration at this point \cite{BR0}.

Since the emitter state Wigner function $f_W^e$ has the spatial 
extent of the
order of the magnetic length $l_B \sim \sqrt{\hbar}$, and exponentially
decays away from this interval, one can expand the scaled action $S/\hbar$ 
in the 
powers of the deviation $(y - y_i) \sim \sqrt{\hbar}$ from the 
``injection'' 
point $y_i$ (defined as the center
of $f_W$) and keep only the terms up to the second order. The integration
over $y$ will then yield:
\begin{eqnarray}
W_{\rm osc} & = & \frac{8}{{m^*}} \Re \sum_\alpha 
{\cal A}_\alpha
\sqrt{
\left|
\frac{ (p^\alpha_i)_z (p^\alpha_f)_z}
     { m_{11}^\alpha + m_{22}^\alpha + 2 }
\right| 
}
\exp\left\{ - \frac{t_\alpha}{\tau_{\rm opt}} - \kappa_\alpha 
\left(\frac{ \Delta p^\alpha_y \ l_B}{\hbar}\right)^2 \right\}
\exp\left( 
i \frac{S^\alpha_{\rm cl}}{\hbar}
- i \frac{\pi n_\alpha}{2} \right)
\label{done_y}
\end{eqnarray}
where the summation is performed over closed orbits which
start and end at the injection point $y = y_i$. Generically there will only be
{\it one} such 
trajectory in each topologically distinct family of closed orbits and this orbit will {\it not}
be periodic, i.e. there will be a non-zero difference between the initial and final momentum.
In Eq. (\ref{done_y}) 
$S^\alpha_{\rm cl}
\equiv
S_\alpha\left(y_i,0;y_i,0\right)$
is the action of this special closed orbit,  
$n_\alpha$ is its topological index\cite{Gutzwiller},
and $\Delta p_\alpha$ is the difference between the ``initial''
and ``final'' momenta:
$\Delta p^\alpha_y = (p_f^\alpha)_y - (p_i^\alpha)_y
$. In (\ref{done_y}) we have 
also introduced the following parameters:
$$ 
\kappa_\alpha = \frac{1}{4 \left(1 + \eta_\alpha^2\right)}, \ \ \
\eta_\alpha = 
\frac{2 - {\rm Tr}\left[ M_\alpha \right]}{2 m_{12}^\alpha}
\frac{l_B^2}{\hbar}, \ \ \ 
{\cal A}_\alpha = 
\sum_{k' k''} {\cal U}_{k' k''} {\cal L}_{k' k''}^\alpha,$$
where the coefficients ${\cal U}_{k' k''}$,  defined in 
(\ref{emitter_wigner}), 
represent the ``transverse'' dependence 
of the emitter state wavefunction, and are non-vanishing only 
for the few lowest Landau levels ($k', k'' \sim 1$).
The general expression for the matrix element ${\cal L}_{k' k''}$ 
is very complicated and will not given here. Here we only
present the result for the most
relevant\cite{Nott,tiltprb,Monteiro} 
families of the closed 
orbits, for which  $\bar{p}^\alpha_y  \equiv 0$:
\begin{eqnarray}
{\cal L}_{k' k''} = \left(\frac{i}{2}\right)^{k' + k''}
\left( 1 + i \eta_\alpha \right)^{ - \frac{k' + 1}{2} } 
\left(1 + i \frac{\xi_\alpha}{\eta_\alpha}\right)^{ - \frac{k'' + 1}{2} }
H_{k'}\left( 
\frac{ \Delta p^\alpha_y l_B}{2 \sqrt{1 + i \eta_\alpha}}
\right)
\ 
\exp\left(
 - i \kappa_\alpha \eta_\alpha 
\left( \frac{ \Delta p^\alpha_y \ l_B}{\hbar}\right)^2 \right)
\end{eqnarray}
where $H_k\left(x\right)$ is the Hermite polynomial, and
$\xi_\alpha = \left(2 - {\rm Tr}\left[ M_\alpha \right]\right) / 
\left(2 + {\rm Tr}\left[ M_\alpha \right]\right)$.

As noted above, typically there is only one trajectory which reaches $y_i$ out of each family of closed orbits and hence only one per family
contributes to the tunneling rate in Eq. (\ref{done_y}). Let us call this orbit the
{\it injected orbit}.  Although
the number of injected orbits 
increases rapidly with the increase of the orbit length,
the contributions of long orbits with the traversal time $t_\alpha$ longer
than the relaxation time $\tau_{\rm opt}$ are suppressed by the damping
factor $\exp\left(-t_\alpha/\tau_{\rm opt}\right)$. Since  $\tau_{\rm opt}$
generally corresponds to the time of a few traversals of the quantum well,
it follows then that one may neglect all but the few shortest injected orbits and 
the sum in (\ref{done_y})  is convenient for comparisons with experimental or numerical data.

It is worth emphasizing again that the trajectories which contribute to the Closed Orbit
Formula (\ref{done_y}) are typically not periodic. The Closed 
Orbit Formula (COF) is therefore free from the problems 
which are encountered in semiclassical periodic orbit 
sums near bifurcations when quadratic expansions fail. 
We present comparisons of this formula with exact numerical results below
in Section IV.B, after outlining our theory of the classical mechanics in the
tilted well.  These comparisons
indicate that the COF yields an excellent quantitative
description of the ghost effect, which is important near bifurcations
in this system \cite{ghost_ref,Monteiro}.
The COF therefore represents a convenient alternative to the use
of uniform approximations\cite{Schromerus} 
and/or complex classical dynamics \cite{Monteiro2}.
The Closed Orbit Formula, Eq. (\ref{done_y}) is one of the main results of the 
current work.

\subsection{Periodic Orbit Formula}

As follows from (\ref{done_y}), only injected orbits with the 
difference between
initial and final momenta $ \Delta p_y \sim \sqrt{\hbar}$, can give
substantial contributions to the tunneling current. Therefore for these important 
injected orbits, in the generic case
(see Fig. \ref{fig:dpy_dy} )  
there exists one and only one periodic orbit (which has $\Delta p_y = 0$) 
in a semiclassically small neighborhood near it.
Using this proximity of each relevant injection orbit 
to a periodic orbit we can re-express the
actions and momenta of the injection orbits in terms of the properties
of their
periodic neighbors (labeled by the index $\mu$)  as follows:
\begin{eqnarray}
S_\alpha\left(y_i, 0; y_i, 0\right) & \simeq & S_\mu 
+ \frac{ {\rm Tr}\left[M_\mu\right] - 2}{2 m_{12}^\mu}
\left(y_i - y_\mu\right)^2, \label{eq:S:clo-po} \\
\bar{p}_y^\alpha & \simeq & p^\mu_y + 
\frac{m^\mu_{11} - m^\mu_{22}}{2 m_{12}^\mu} \left(y_i - y_\mu\right)
\label{eq:p:clo-po}
\end{eqnarray}
and for the tunneling current current we obtain
\begin{eqnarray}
W_{\rm osc} & = &\frac{16}{{m^*}}
\int dy \int dp_y
\ f_{W}^{(e)}\left(y,p_y\right)
\ \sum_{\mu}
\frac{ p^{\mu}_z \exp\left( - \frac{T_\mu}{\tau_{\rm opt}} \right) }
{\sqrt{|m_{11}^{\mu} + m_{22}^{\mu} + 2|}}
\ \cos\left[
\frac{S_\mu}{\hbar} - \frac{\pi n_\mu}{2}
+  Q_{\mu}\left( \delta y,\delta p_y\right)
\right]
\label{w_osc}
\end{eqnarray}
where
\begin{eqnarray}
Q_{\mu} & = &
\frac{2}{\hbar}
\frac
{
m_{21}^{\mu} \left(\delta y\right)^2
+
\left(m_{22}^{\mu} - m_{11}^{\mu}\right)
\delta y \delta p_y
-
m_{12}^{\mu}
\left(\delta p_y\right)^2
}
{
\left( m_{11}^{\mu} + m_{22}^{\mu} + 2 \right)
}. \nonumber
\end{eqnarray}
Here $\mu$ labels the periodic orbit, $\delta y = y - y_{\mu}$,
$\delta p_y = p_y - \left( p_\mu \right)_y$,
the integer $n_\mu$ is
the topological index \cite{Gutzwiller} of the periodic orbit,
and the
$2\times2$ monodromy matrix\cite{Gutzwiller} $M = (m^{\mu}_{ij})$  is
calculated at contact point at the emitter barrier.

This relation between injected orbits and periodic orbits is 
generically valid because if 
$\Delta p_y$ is small for the injected orbit, then the function 
$\Delta p_y (y)$ for this
family of closed orbits can be expanded linearly around $y_i$ and 
extrapolated to zero to
find the nearby periodic orbit (see Fig. \ref{fig:dpy_dy}(a) ).  
However if that periodic orbit is in the 
neighborhood of a bifurcation this expansion will fail and 
Eq. (\ref{w_osc}) will become
unreliable.  The simple case of a tangent bifurcation is illustrated 
schematically in Fig. \ref{fig:dpy_dy} (b).
Here just before the bifurcation there are two periodic orbits which 
are close together, neither of which alone can adequately describe the 
contribution of the relevant closed orbits.  
Just after the tangent bifurcation there is no periodic orbit at all in the 
semiclassically small neighborhood of the injected orbit.  There may 
however be another 
periodic orbit of the same topology far away from the injected orbit - 
see Fig. \ref{fig:dpy_dy}(c). 
In this case the linear expansion will yield a semiclassically large 
value for $\Delta p_y$ and it will not contribute to the sum, 
introducing no additional error,
but missing higher-order contributions near the bifurcation.  
In such cases the periodic orbit formula (\ref{w_osc}) will lose accuracy 
but not fail completely as the resulting divergences
are integrable (see Fig. \ref{fig:BVplots} below).  However for such 
cases the Closed Orbit Formula will remain quite precise and can be adapted 
for the specific type of nearby bifurcation.  In section IV.B we 
demonstrate this 
for the case of tangent bifurcations in which we calculate the exponentially
 small ``ghost'' contributions \cite{ghost_ref} beyond the bifurcation.

Eq. (\ref{w_osc}) has the feature that it separates out the 
semiclassical summation over periodic orbits within the well from the 
contribution of the injection function $f_{W}^{(e)}\left(y,p_y\right)$ which 
cannot be calculated semiclassically and 
depends
on the doping level in the emitter, as well as the magnetic field and 
tilt angle, thus it
is convenient for a general quantitative description of the actual experiments.
If the emitter state wavefunction has only zero-order Landau level
component (${\cal L}_{k' k''} \sim \delta_{k'0} \ \delta_{k''0})$, then
the integration in (\ref{w_osc}) can be performed analytically. For the
most relevant periodic orbits with $p_y = 0$ at the point of collision
with the emitter barrier, we obtain:
\begin{eqnarray}
W_{\rm osc} & = &\frac{8 \sqrt{2}}{{m^*}}
\ \Re \ \sum_{\mu} \ 
\frac
{ 
p^{\mu}_z 
\exp\left( - \frac{T_\mu}{\tau_{\rm opt}} - R_\mu\right) 
}
{
\sqrt{{\rm Tr}\left[M_\mu\right] + 2 i 
\left( \frac{m_{21}^\mu}{eB\cos\theta} - 
m_{12}^\mu eB\cos\theta \right)}
}
\ \exp\left[i 
\frac{S_\mu + \Delta S_\mu}{\hbar} - i 
\frac{\pi \tilde{n}_\mu}{2}
\right]
\label{br}
\end{eqnarray}
where
\begin{eqnarray}
\Delta S_\mu = 
\frac{\sigma_\mu}{1 + \sigma_\mu^2} \ eB\cos\theta 
\left(y_\mu - y_i\right)^2, \ \ \ 
R_\mu = \sigma_\mu \ \Delta S_\mu, \ \ \
\sigma_\mu = 
\frac{ {\rm Tr}\left[M_\mu\right] - 2}{2 m_{12}^\mu}
\frac{1}{eB\cos\theta}, \ \ \
\tilde{n}_\mu = n_\mu + 1 - 
{\rm Sign}\left({\rm Tr}\left[M_\mu\right] + 2\right).
\nonumber
\end{eqnarray}
If the shift of the ``injection point'' $y_i$ from zero can be 
neglected, Eq. (\ref{br}) reduces\cite{Saraga} to the result, 
obtained by Bogomolny and Rouben\cite{BR}.

The summation in (\ref{w_osc}) is performed over all isolated
periodic orbits, both stable and unstable.  Near stable islands
the motion is regular and we expect semiclassical quantization
to yield discrete energy levels (neglecting phonon broadening)
and sequences of eigenfunctions
localized on the islands  \cite{Miller,Monteiro}.
In contrast near unstable orbits
the motion is chaotic and semiclassical theory does
not yield discrete levels \cite{Gutzwiller}.  This difference,
can be displayed explicitly by performing exactly the
summation over repetitions of the primitive periodic orbits
in Eq. (\ref{w_osc}), yielding
\begin{eqnarray}
W & = & \frac{8}{m} \sum_\mu \left(p_\mu\right)_z \sum_\ell
\Delta\left(\frac{T_\mu}{\tau_{\rm eff}^\mu},
\frac
{
S_\mu\left(\varepsilon_\ell\right)
}
{\hbar}
- \frac{\pi n_\mu}{2}
\right)
\int dy \int dp_y 
\ 
f_W^e \left(y, p_y\right)
\
g_\ell^{\mu,\pm} \left(y, p_y\right)
\label{w_po}
\end{eqnarray}
where
$\Delta\left(\sigma,\rho\right) =
\frac{\sinh\left(\sigma\right)}
{\cosh\left(\sigma\right) - \cos\left(\rho\right)}$,
and the index $+$ or $-$ denotes stable or unstable orbits.
The quantity $\hbar/\tau_{\rm eff}^\mu$ is an effective level-broadening
which differs in the two cases.

Note that the function $\Delta$ has a peak
every time the semiclassical quantization condition
$S_{\mu}(\varepsilon_l)= 2 \pi \hbar (n + n_{\mu}/4)$
is satisfied, and these peaks become delta functions as
$\tau_{\rm eff}^\mu \to \infty$.  For stable orbits
we find that $\tau_{\rm eff}^\mu = \tau_{\rm opt}$, so if we neglect
phonon scattering ($ \tau_{\rm opt} \to \infty$) we do recover
perfectly discrete contributions to the tunneling current.
In the stable case the argument of $S_{\mu}$
in Eq. (\ref{w_po}) is $\varepsilon_\ell^+ \equiv
\varepsilon - \hbar \omega_\perp^{\mu,+} (\ell + 1/2)$ which may
be interpreted as the energy of longitudinal motion along the orbit.
Due to the harmonic approximation
the quantization of the transverse oscillations around the periodic orbit
simply yields equally spaced levels \cite{Monteiro,Miller}
with spacing $\hbar \omega_\perp^{\mu,+}$,
where the frequency $\omega_\perp \equiv \phi_\mu/T_\mu$, 
$\phi_{\mu}$ is the winding number  \cite{Gutzwiller} and $T_{\mu}$ the 
period
of the orbit.  So the discrete energies at which tunneling occurs
{\it are} the correct semiclassical energy levels of the well.

The amplitude of each contribution is given by the
coefficient functions $g_{\ell}$ in Eq.
(\ref{w_po}), which are the Wigner transforms of the harmonic oscillator
wavefunctions corresponding to these transverse modes:
\begin{eqnarray}
g_\ell^{\mu,+}\left(y, p_y\right) & = &
(-1)^\ell
L_\ell\left(2 \left|\tilde{Q}_{\mu}\right|\right)
\exp\left(- \left|\tilde{Q}_{\mu}\right| \right)
\label{w_stable}
\end{eqnarray}
where $L_\ell$ is the Laguerre polynomial and 
$
\tilde{Q}_{\mu} = 
\left| 2 + {\rm Tr}\left[M\right] \right|^{1/2}
\left| 2 - {\rm Tr}\left[M\right] \right|^{-1/2}
Q_\mu $.

Since the result (\ref{w_stable}) is
based on the harmonic approximation within a stable island,
we may only include modes up to $\ell_{\rm max}$, which
is given by the ratio of the island area to $\hbar$.
Phonon scattering smears out each of these discrete contributions
to $W_{\rm osc}$ over an energy range $\hbar/\tau_{opt}$.

In contrast, for unstable periodic orbits
we find that
$\tau_{\rm eff}^\mu = \tau_{\rm opt}/
\left(1 + \ell \lambda_\mu \tau_{\rm opt} \right)$, where $\lambda_\mu$
is the Lyapunov exponent near the orbit $\mu$. Hence this time is finite
and equal to $1/\lambda_{\mu}$ in the absence of phonon scattering.
Therefore, instability acts as a sort of intrinsic level-broadening,
and each PO describes a contribution due to a cluster of levels.
The peak of the function $\Delta$ in Eq. (\ref{w_po}) corresponding to the
mean energy of the cluster is given by
$S_{\mu}(\varepsilon)= 2 \pi \hbar (n + n_{\mu}/4)$. The weight
functions
\begin{eqnarray}
g_\ell^{\mu,-}\left(y, p_y\right) & = &
(-1)^\ell
\Re\left\{
L_\ell\left(2 i \tilde{Q}_{\mu}\right)
\exp\left( i \tilde{Q}_{\mu} \right)
\left(1
+ i
\frac
{\sin\left(S_\mu/\hbar - \pi n_\mu/2 \right)}
{\sinh\left( T_\mu / \tau_{\rm eff}^\mu \right)}
\right) \right\}
\label{w_unstable}
\end{eqnarray}
are related to Wigner functions
{\it averaged} \cite{Berry} over the eigenstates of the cluster.
For each unstable PO the high-$\ell$ contributions are
exponentially damped, and the main contribution to the
tunneling rate is given by the $\ell=0$ term.

We now have a rigorous criterion for which periodic orbits contribute
substantially to the tunneling current in Eq. (\ref{w_po}).
The injection function $f_W(y,p_y)$ is centered on $y_i$ and $p_y=0$ with
widths $\sim l_B,\hbar/l_B$ in $y,p_y$.
From Eqs. \ref{w_stable},\ref{w_unstable} and the definition of
$\tilde{Q}_{\mu}$ we find that classical weight functions $g_{\mu}$ are
are centered at $y_{\mu}, (p_y)_{\mu}$
with widths $\sim  l_\mu  =  \sqrt{2 \hbar \left|m_{12}^{\mu}\right| }
\left| 4 - {\rm Tr}^2\left[M_\mu\right] \right|^{-1/4} $ and
$\hbar/l_{\mu}$ respectively.   When the real and momentum space
peaks of these two functions overlap the PO is semiclassically
``accessible'' and makes a substantial contribution to the tunneling
current.\cite{bifurcations}

\section{Classical Dynamics}

\subsection{Model Hamiltonian}

The semiclassical formula for the tunneling current, Eq. (\ref{w_po}), 
involves 
only classical periodic orbits within the well at energies below the top 
of the barriers.  Therefore we can model the system by taking the barriers 
to be infinite perfectly-reflecting walls at $z=0, d$.  With our choice of 
gauge and coordinates we have 
\begin{eqnarray}
H & = & \frac{(p_x - e B y \cos(\theta) + e B z \sin(\theta))^2}{2{m^*}} 
+ \frac{p_y^2}{2{m^*}} + \frac{p_z^2}{2{m^*}}  + eE z \nonumber \\  
&+& U(-z) + U(z - d) \label{h3d} 
\end{eqnarray} 
where the functions $U$  ( $U(z<0) = 0, \ U(z>0) = \infty $ ) represent 
the infinite walls.  Because $H$ is 
independent of x, and $p_x$ can be eliminated by a gauge transformation 
\cite{ss,Bell}, this problem can be simplified to a two-dimensional 
effective hamiltonian \cite{tiltprb}. The original hamiltonian involves 
four variable experimental parameters: $B,E,\theta,d$ and, as in the 
case of 
diamagnetic hydrogen, it is essential to express H in terms of the minimum 
number of dimensionless parameters. There are only two energy-independent 
classical length scales in the problem: the barrier spacing, $d$, and 
$l_D = (E/B)T_c$, where $E/B$ is the drift velocity for perpendicular 
fields and 
$T_c$ is the cyclotron rotation period ($T_c=2\pi/\omega_c$).  Since the 
length-scale d will drop out of the problem when the electron energy 
$\varepsilon < eEd=eV$ it is natural to scale all lengths by $l_D$.  When 
this is done \cite{tiltprb} the effective hamiltonian is found to depend 
on three dimensionless parameters\cite{scalings}: 
$\theta,\beta = 2v_0B/E$ and $\gamma = 
\varepsilon/eV$, where $v_0$ is the velocity corresponding to the total 
injection energy $\varepsilon =  m^* v_0^2/2 = \varepsilon_i + eV$, where 
$\varepsilon_i$ is the emitter state injection energy into the well.  This 
energy is typically small compared to $eV$.  Equally importantly, its 
ratio to the voltage drop is roughly constant in experimental conditions, 
implying $\gamma \approx 1.15$ and constant.  Hence for fixed tilt angle the 
scaled dynamics need only be studied as a function of one parameter, 
$\beta$, which can be regarded as the scaled magnetic field.  

\subsection{Poincar\'e Map and the Origin of Chaos}

The most convenient surface of section to choose is the plane $z=d$ 
corresponding to the collector barrier, as all trajectories within the 
well impinge on this plane (although in certain cases the emitter SOS may 
be useful as well).  The most convenient phase-space coordinates for the 
system are the scaled velocities $v_x/v_0,v_y/v_0$ (we note that for this 
system $v_x \sim y$).  
When $\theta=0$ the (integrable) motion consists of cyclotron rotation 
perpendicular to the E field and uniform acceleration along it 
(perpendicular to the barriers).  Collisions reverse the $z$-component of 
velocity and conserve separately longitudinal and cyclotron energy.  When 
$\theta \neq 0$ the motion between collisions is still integrable, 
although now there is a drift velocity added to the cyclotron motion and 
the acceleration is along the magnetic field with magnitude $\sim E \cos 
\theta$.  However collisions now exchange energy between the cyclotron + 
drift energy and the longitudinal kinetic energy.  Since the tilt angle is 
in the $y-z$ plane it is straightforward to see  that there is no 
energy exchange at collisions when $v_y=0$ and maximal energy exchange 
when $v_x=0$.  This then is the origin of chaos in the system, the energy 
exchange between the two degrees of freedom depends on the phase of the 
cyclotron rotation at each collision, very much as in the Chirikov 
standard map (or kicked rotor)\cite{Chirikov}, and just as in that 
case the system becomes chaotic
as this frequency increases relative to the collision frequency (for 
non-zero tilt angle) - see Fig. \ref{fig:origin_of_chaos}.   
Hence as B increases or V decreases the system 
undergoes a transition to global chaos. A simple scaling analysis shows 
that for fixed $\gamma$ the classical parameter $\beta \sim B/\sqrt{V}$, 
leading to parabolic curves of constant classical dynamics in the $B-V$ 
plane \cite{tiltprb}.

In order to derive a simple criterion for the transition to chaos in this 
system Shepelyansky and Stone \cite{ss} noted that, since $\gamma$ was 
close to unity, many trajectories would have insufficient longitudinal 
energy to reach to emitter barrier and one could consider simply electron 
motion in the triangular well formed by the electric field and the 
collector barrier (the ``single-barrier'' model).  In this approximation 
they showed that the dynamics reduces to an effective standard map with a 
smooth KAM transition to chaos and a chaos parameter given by $K = 2 
\theta \beta $.  When $K \sim 1$ the last KAM curve of the system 
breaks and one has full exchange of energy between the degrees of freedom. 
 However our recent work has shown that this approximation does not 
capture many crucial and interesting aspects of the physics, since those 
trajectories which {\it do} reach the emitter barrier are crucial, and including 
the second barrier changes qualitatively the dynamics in that region of 
phase space.  The single-barrier model is a mathematically tractable and 
experimentally realizable KAM system; it is studied in detail in ref. 
\cite{tiltprb}.  Here we will only consider the double-barrier model 
relevant to current experiments and to our semiclassical tunneling formula 
above.

\subsection{Non-analytic Dynamics in the Double-Barrier Model}

The key new feature of the double-barrier model is that the presence of 
the second barrier introduces non-analytic behavior into the Poincar\'e 
map.  This map takes the values of $v_x,v_y$ at a collision and maps them 
to the values at the subsequent collisions.  In the double-barrier model 
the map defines a closed curve in the SOS corresponding to those initial 
conditions which reach the second barrier with $v_z=0$ exactly.  The electron
with the initial velocity inside 
this ``critical boundary'' reach the second 
barrier and get an additional ``kick'', and otherwise it will not.  
It is easy to show then that the map will have discontinuous derivatives 
on this boundary.  As a result, just as in the stadium billiard or other 
systems described by non-analytic maps, the system in the vicinity of the 
critical boundary undergoes a sudden jump to chaos instead of a smooth KAM 
transition.  However since trajectories sufficiently far from the critical 
boundary are confined away from the boundary by KAM curves, unlike the 
stadium one does not have global chaos for arbitrary tilt angle, instead 
one has a ``chaotic halo'' region surrounding the critical boundary (see 
Fig. \ref{fig:halo}).  This region becomes of substantial size well before 
the phase 
space becomes globally chaotic and hence has a significant impact on the 
transition to chaos.  We will see below that all of the periodic orbits 
relevant to the semiclassical tunneling formula are born precisely at the 
critical boundary in the chaotic halo region.  They are born in a novel 
type of bifurcation we call a ``cusp bifurcation'' which we will discuss 
in detail below.  These bifurcations were very difficult to detect 
numerically from the SOS because they violate the bifurcation theorems for 
analytic hamiltonian maps.  Understanding of the short periodic orbits and 
their bifurcations came from continuity arguments from the $\theta=0$ 
limit and from analytic results to be discussed below.

\subsection{Periodic Orbits at $\theta = 0$}

All periodic orbits can be easily classified at $\theta=0$.  First, there 
is a single isolate ``traversing orbit'' (TO) which has zero cyclotron 
energy and follows the line normal to the barriers, bouncing back and 
forth. Then there are an infinite number of families of ``helical orbits'' 
(HO's) with periods $T$ given by the resonances between the cyclotron 
rotation with period $T_c$ and the longitudinal bouncing-ball motion with 
period $T_L$:
\begin{eqnarray} 
T = n T_L = k T_c. 
\end{eqnarray}
In the collector Poincar\'e map this would define a family of period-$n$ 
orbits, all related to one another by rotation of the helix.
An important point is that when $\gamma > 1 $ (the case we are interested 
in) there are can be two such families (or ``rational tori'') for each 
pair of integers $(n,k)$.  For simplicity assume that $V$ and $\varepsilon_0$ 
are fixed and we are varying B.  The resonance condition can then be 
rewritten as
\begin{eqnarray} 
T_L= \frac{2 \pi {m^*} k}{eB n} 
\end{eqnarray}
where $T_L$ is a function only of the fraction of the total energy which 
is in the longitudinal motion, $\varepsilon_L$.  The dependence of $T_L$ on 
$\varepsilon_L$ is shown in Fig. \ref{fig:tl_e}.  When $\varepsilon_L < eV$ 
then $T_L$ 
increases as $\sqrt{\varepsilon_L}$; however when $\varepsilon_L > eV$ then 
$T_L$ begins to decrease because the electron reaches the emitter barrier 
and the bounce returns it faster to the collector barrier.  Now consider, 
for any given resonance $(n,k)$, increasing $B$ from zero.  Until 
$B_1 \equiv 
\pi \left(  {m^*} E\right)^{1/2} \left(2 e d\right)^{-1/2} k/n$ 
there are no solutions for the resonance condition, meaning that the 
cyclotron rotation is too slow for that particular resonance to occur (here we
have used the explicit form of $T_L(E,d)$).  For 
$B_1 < B < 
B_2 \equiv B_1 \left(\sqrt{\gamma} + \sqrt{\gamma-1}\right)$ 
there are {\it two} solutions corresponding to two 
different families of helical orbits with the same $T_L$; one of which has 
longitudinal energy greater than $eV$ and reaches the emitter barrier, the 
other of which has $\varepsilon_L < eV$ and doesn't reach the second barrier 
(see Fig. \ref{fig:tl_e}).  We will refer to these two types henceforth 
as ``emitter'' and 
``collector'' orbits.  As B is increased further the emitter orbit must {\it 
give up} cyclotron energy to remain in resonance, while the collector 
orbit must {\it gain} cyclotron energy in order to remain in resonance.  
Eventually at $B=B_2$ the emitter orbit has zero cyclotron energy and is 
degenerate with the traversing orbit, which exists at all magnetic fields 
for each energy.   For $B > B_2$ the emitter family of periodic orbits 
ceases to exist, but the collector family persists for all $B > B_1$.  

In the surface of section this evolution look as follows (Fig. 
\ref{fig:schem_psos}): at 
$B_1$ a circle of degenerate period-$n$ orbits appears exactly at the 
critical boundary (i.e. reaching the emitter barrier with $v_z=0$.  As B 
increases this splits into two circles, one with radius greater than the 
critical boundary (the collector torus) and one with radius smaller (the 
emitter torus).  As B tends to $B_1$ the radius of the inner circle 
shrinks to zero and the torus disappears.  The conclusion is that at 
$\theta = 0$ all periodic emitter orbits (except the traversing orbit) 
exist only for a finite interval of magnetic field (and a similar 
conclusion holds if $V$ is varied as well).  By continuity the same must 
be true for $\theta \neq 0$.

\subsection{Effect of Non-zero Tilt on Periodic Orbits}

Tilting the magnetic field makes the dynamics non-integrable and as usual 
all families of periodic orbits are destroyed, being replaced by an equal 
number of stable and unstable isolated fixed points with the same 
periodicity in the map.  Isolated periodic orbits (in this case the 
traversing orbit) survive the perturbation, and the resonances of these 
orbits transform into its bifurcations.  In ref. \cite{tiltprb} we have 
given an exhaustive analysis of the classical periodic orbit theory in 
this system.  Here we will simply consider briefly the case of period-two 
orbits which is both illustrative of the general situation and most 
relevant to the experiments.  In particular we will consider the 
period-two orbits, which occur at the 
lowest values of magnetic field (these orbits at $\theta=0$  make 1/2 a 
cyclotron rotation per collision with the emitter).  There are four such 
orbits, with different topology, the $(0,2)_\Lambda$,$(2,2)_\Lambda$, and 
$(0,2)_V$,  $(2,2)_V$ orbits (see Fig. \ref{fig:po2}). The first and second 
numbers in this notation
represent the numbers of bounces per period with the emitter and 
with the collector respectively, and the subscript describes the ``topology''
of the orbit.

Since the emitter orbits have different topology they cannot be created 
together e.g. the $(2,2)_\Lambda$ must be created with the 
$(0,2)_\Lambda$ and the $(2,2)_V$ 
with the $(0,2)_V$.  Since $(0,2)_\Lambda$,$(0,2)_V$ are collector orbits and 
$(2,2)_\Lambda$,$(2,2)_V$ 
are emitter orbits, this can only happen precisely at the critical 
boundary (just as for the unperturbed case); the bifurcation at which they 
are born has the non-analytic properties mentioned above which lead to a 
cusp at the bifurcation point in the bifurcation diagram
as shown in Fig. \ref{fig:bif_cascade}.  In correspondence to the 
evolution of tori at $\theta = 0$ shown in 
Fig. \ref{fig:schem_psos}, now as  $\beta$ (or magnetic field) is 
increased the emitter orbits fixed points move inward in the SOS (but now at 
slightly different rates for the two different emitter orbits), eventually 
annihilating with the period-one 
traversing orbit in successive backwards period-doubling bifurcations.  This 
scenario is general for all emitter orbits: birth at the critical boundary 
in a cusp bifurcation; death in a backwards bifurcation in the neighborhood 
of the traversing orbit \cite{period-three}.

There is an important sub-plot in this scenario.  Once the symmetry of the 
problem has been broken by non-zero tilt angle it is not possible for a 
$(2,2)_V$ 
and $(0,2)_V$ orbit to be born together, because the two ``arms'' of the $V$ 
orbit can no longer be equal in length.  Hence the two $(1,2)$ orbits are 
born first, one of which has an arm reaching the emitter barrier and one 
of which does not.  With increasing $\beta$ the emitter orbit evolves into 
a $(2,2)_V$ orbit, typically not directly, but through a bifurcation 
``cascade'' \cite{explain_primary} of 
the type shown in Fig. \ref{fig:bif_cascade}.  In such a cascade a $(1,2)$
 orbit born
in a cusp bifurcation with a $(0,2)$ will disappear in a tangent bifurcation 
with another $(1,2)$ orbit born in a cusp bifurcation with a $(2,2)$.  Thus 
the first $(1,2)$ orbit is
not directly connected by bifurcation to the $(2,2)$, but only indirectly as 
shown in
Fig. \ref{fig:bif_cascade}.

Such bifurcation cascades connected to a resonance of the traversing orbit 
are a generic feature of this system, for period-$N$ orbits. The only 
difference is that for the orbits with many bounces in 
the corresponding bifurcation cascade the ``connection'' of the $(0,N)$
and $(N,N)$ orbits, required by continuity to the behavior at zero tilt angle,
is a ``multi-step process'', which includes all the 
``intermediate'' periodic orbits of the sequence $(1,N)$, $\ldots$, $(N-1,N)$.
Thus, despite the increasing complexity, all
bifurcation cascades in this system can be analyzed within a single framework.

Each of these bifurcation cascades terminates finally in a
resonance with the period-one traversing orbit. The 
analytic expressions for the 
period and stability of all period-one orbits, derived in Ref. 
\cite{tiltprb}, provide a
straightforward description of each of these resonances, and therefore 
allow a direct and unambiguous identification of all of the 
primary bifurcation cascades.

\subsection{Cusp Bifurcations and Stability Analysis}

The existence of both emitter and collector period-one helical orbits
at $\theta = 0$, which appear together at the critical
boundary,  implies that for $\theta \neq 0$ their 
counterparts must be produced in cusp bifurcations. The analytic
analysis of the period-one orbits allowed us to obtain a 
detailed description of these cusp
bifurcations. For example we were able to show that these orbits do not 
tend to marginal 
stability at the cusp bifurcation, in contrast to the behavior required of 
analytic maps.  This led to some more general results on the properties of 
all CB's, which are as follows: 1) all cusp bifurcations appear at the 
critical boundary, where the Poincar\'e map is non-analytic; 2) any cusp 
bifurcation produces two periodic
orbits, which differ by one ``point'' of contact with the emitter wall;
depending on the topology of these orbits, they differ either by 
one or by two bounces with the emitter barrier per period; 3) out of these
two new orbits, the one which has more bounces with the emitter wall
(the ``more connected'' orbit), is ``infinitely unstable'' 
(${\rm Tr}[M] \to \infty$), while it's partner might
either be stable or unstable. 

The calculation of the trace of the monodromy
matrix for the $V$-type period-$2$ orbits, presented in the 
Fig. \ref{fig:trM} , 
illustrates the latter point. Note the divergence of the ${\rm Tr}[M]$ for
the orbits $(1,2)^+$ (the ``more connected'' partner of $(0,2)_V$) and 
$(2,2)_V$ (the ``more connected'' partner of $(1,2)^-$). Meanwhile, 
their partners appear either stable ($(1,2)^-$) or unstable ($(0,2)_V$).

We hypothesize that the presence of cusp bifurcations with these properties is
a general feature of hamiltonian maps with hard wall potentials when the orbits
can either hit or miss one part of the hard wall. 
 Therefore the same phenomenology
should exist in certain billiard systems (e.g. the ``annular 
billiard''\cite{annular}), or in billiards in strong magnetic field 
\cite{blaschke}, the possibilities we are currently 
investigating
\cite{annular_unpub}.  This phenomenology has importance for the quantum
properties of the system, because it leads to special `metastable orbits'' 
as we now
discuss in the case of the period-two orbits in the tilted well.

As the effective chaos parameter $\beta$ increases, 
the collector orbit $(0,2)_V$
becomes increasingly more unstable, as expected for a periodic orbit of a
system undergoing the transition from integrability to chaos. However,
the other ``least connected'' orbit of this bifurcation cascade, the 
orbit $(1,2)^-$, shows a dramatically different behavior. Initially,
with the increase of $\beta$ it loses stability, as expected. However, 
this orbit will later have to annihilate with the orbit $(1,2)^+$ in a 
backwards tangent bifurcation. Since the tangent bifurcation can only involve
two marginally stable orbits, the initially stable $(1,2)^-$ can not 
develop strong unstability and must stay either stable or only marginally 
unstable in it's whole interval of existence.

Note, that this continuity argument applies to any ``intermediate''
period orbit, which appears as the ``less connected'' 
partner in a cusp bifurcation, and annihilates in a tangent bifurcation.
Therefore, there must be at least one such metastable orbit in any 
primary bifurcation cascade of this system. This ``metastability'' of a 
certain subset of the classical periodic orbits
in the quantum well in a tilted magnetic field has profound consequences
in its quantum spectrum.

\subsection{The Origin of Strong Scarring}

It is now well known that certain
quantum wavefunctions of classically chaotic systems can strongly
localize near 
unstable periodic orbits. Such wavefunctions are  termed ``scars'' and 
their properties have been extensively studied and well understood
during the last decades\cite{Heller,Agam,Fishman,Kaplan}. 
The scarred wavefunction are most 
likely to be associated with the periodic orbits which are not too unstable. 
A sufficient
condition of the strong scarring is $\lambda T \lesssim 1$, where $\lambda$
is the instability (Lyapunov) exponent associated with the orbit, and $T$ is 
it's period. Typically, in chaotic systems, the periodic orbits quickly
become increasingly more unstable, as the classical parameter driving the 
system to chaos, is increased. Such orbits can strongly scar the
wavefunctions only in a small interval of classical parameter space
in which they are not too far from stability. 

Therefore, it came as a surprise, when a 
persistent scarring by the same orbit in a large interval of applied voltage,
was observed both numerically\cite{Fromhold_scars_prl} and (indirectly) 
experimentally\cite{Fromhold_scars_nature}. Although it has been argued
\cite{comment} (with some merit in our view)
that the experimental results are in a regime of low quantum numbers 
for which the
concept of scarring becomes  ambiguous, extensive quantum numerical 
calculations
confirm a strong periodic scarring at high quantum numbers as well.
This phenomenon is naturally explained
by the ``metastability'' of the ``intermediate'' periodic
orbits in this system. Since these orbits
remain close to stability over their whole interval of existence, they
produce strong scars in the quantum wavefunctions in a large interval
of the applied voltage or magnetic field. This behavior is illustrated in 
Fig. \ref{fig:scar_strength}, which
represents the calculation of the scar
``strength'' (defined as the value of the Husimi projection of the
quantum wavefunction, taken at the 
fixed point the periodic orbit\cite{dss}) 
due to the $(1,2)^-$ orbit.

As we already pointed out, there is at least one ``metastable'' orbit in each
of the 
primary bifurcation cascades. The semiclassical theory therefore implies
that each of these orbits should also produce strong scars in quantum 
wavefunctions. Indeed, a detailed analysis\cite{tiltscar} of the wavefunctions
in the quantum well uncovers strong and persistent scarring due to 
the ``metastable'' period-$3$ orbit $(1,3)$ and period-$5$ orbit $(1,5)$.

\subsection{``Exchange Interaction'' between Bifurcation Cascades}

An interesting feature of the classical dynamics in the quantum well in a 
tilted magnetic field, noted in Ref.\cite{tiltscar}, is the exchange
bifurcations between the orbits of the different primary bifurcation
cascades. Fig. \ref{fig:exchange_cascade} 
depicts the exchange bifurcation between $V$-type periodic
orbits of two different period-$2$ cascades. For convenience of the further
analysis, we represent the relevant orbits by the $x$-components of the 
velocity $v_x \propto y$ at the {\it emitter} barrier (by definition, 
such diagrams do 
not show the collector orbits). As the tilt angle approaches the 
critical value of $\theta^* \approx 30^\circ$, the orbits $(1,2)^-$ and 
$(1,2)^+$ of the first
cascade, and the orbit $(1,2)^+_*$ of the second cascade come close to each
other, and at the critical angle exchange partners, so that for $\theta
> \theta^*$ the orbit $(1,2)^+_*$ is now paired with 
$(1,2)^-$ and belongs to the first cascade, 
whereas the orbit $(1,2)^+$ is paired with $(1,2)^-_*$ and belongs 
to the second cascade. As the tilt angle is increased further,
the two reconfigured bifurcation cascades move away from each other - see 
Fig. \ref{fig:exchange_cascade}c.

\section{Semiclassical Tunneling Spectra and 
Comparison with Experiment and Numerical Calculations}

\subsection{Periodic Orbit Theory}

Since the motion of the fixed points
of the relevant periodic orbits near the exchange bifurcation
strongly affects their accessibility for tunneling,
the exchange bifurcation dramatically
manifests itself in the tunneling spectra. This is demonstrated in Fig. 
\ref{fig:exchange_cascade},
where the gray-scale regions (calculated at $B=8T$) represent the 
semiclassical width $l_{\mu}$ of these wavefunctions localized near 
the corresponding periodic orbits.
The width of the injection function
$f_W(y)$ is represented by the region between dashed lines.
Whenever these regions overlap for some value of $\beta$
the orbit is accessible and Eq. (\ref{w_po}) predicts
that a peak-doubling region will appear in the $B-V$ parameter space
along the parabola corresponding to that value of $\beta$ 
(see Fig. \ref{fig:BVplots}) \cite{explain_why_py_irrelevant}.

At $\theta=29^{\circ}$ the accessibility intervals for the $(1,2)^-$
orbit and the $(1,2)^+_*$ orbit cover the entire interval 
$4.3 < \beta <10.9$
overlapping briefly around $\beta=7.5$.  Thus one expects a large
region of peak-doubling in the $B-V$ plane with no gaps as observed
in the experiment\cite{Bell}. Since the periods of the relevant orbits 
are almost
degenerate, these distinct periodic orbits will produce the oscillations
of the tunneling current of almost the same frequency. Note, that the
orbits $(1,2)^-$ and $(1,2)^+$ of the first cascade are slightly better 
coupled with the emitter state, that the orbit $(1,2)^+_*$. 
Therefore, in the
I-V, we expect the amplitude of oscillations to be higher in the 
high-voltage region ($\beta \sim 1/\sqrt{V}$).

As $\theta = 31^\circ$, immediately after the exchange bifurcation, 
the relevant orbits of the newly-formed ``first cascade'' start 
to move away from the accessibility region. As a result, the amplitude
of the oscillations of the tunneling current 
is now higher in the low-voltage
region, and we expect an abrupt amplitude change
around $\beta\approx 7.3$, ($V \approx 0.44$V for $B=8$T).  
This is observed clearly in the experimental data 
of Fig. \ref{fig:BVplots}(a),(d), where the amplitude 
change is compared to theory.  The theory is found to predict the ratio
of the amplitudes between the low and high voltage regions
(above $0.45$V) with only $25\%$ mean error.

At $\theta \approx 34^{\circ}$, the periodic orbits of the first cascade
are no longer accessible. At the same time the
$(1,2)^+$ orbit is still highly accessible at low voltages (high $\beta$).
Thus we expect the high voltage oscillations to disappear at
$\theta= 34^{\circ}$ while the low voltage oscillations persist.
This is seen clearly in Fig. \ref{fig:BVplots},
 again in good qualitative and
quantitative agreement with theory.  

In Fig. \ref{fig:BVplots}, we plot the peak position
data of Ref. \cite{tiltsc} in the entire $B-V$ plane against 
the semiclassical
prediction based on the contributions of these periodic orbits.
Again, the comparison shows a good agreement between semiclassical
calculation and the experimental data \cite{endnote}.

\subsection{Beyond Periodic Orbits: The Closed Orbit Theory}

As was demostrated in the previous section, the periodic theory can 
be successfully used for the quantitative description of the tunneling spectra
over a wide range of system parameters. However, the periodic orbit approach
may fail near the bifurcations, when the relevant orbits are not isolated, and
the second-order expansion, used to obtain the POF, is not justified. A manifestation of 
this can be
seen in Fig. \ref{fig:BVplots}a,d, where one sees a spurious dip in the semiclassical trace
near $V = 0.3V$ which has no analog in the experimental $I-V$ curve.  This dip is precisely due to the occurrence of a bifurcation near this voltage.
Also, the POF provides no description for the ``ghost effect'' near tangent bifurcations, when the contribution of the particular orbits which disappear in the bifurcation remains in 
the spectrum even after their disappearance\cite{ghost_ref}.

Generally, the regions where the POF is inadequate, are 
{\it semiclassically small},
i.e. the corresponding intervals $\Delta V, \Delta B \sim \hbar$.
And in the case of the ghost effect, the ghost contributions decay exponentially
away from the bifurcation point. However, due to a relatively
large value of the effective $\hbar$ in the 
experiment\cite{Nott,Bell,Fromhold_scars_nature} (particularly in Ref. 
\cite{Fromhold_scars_nature}),
the measured tunneling spectra
do show clear features related to these effects. 
In particular, as it has been shown by Monteiro et al.
\cite{Monteiro} that a substantial part of the peak-doubling 
region at $\theta = 27^\circ$ is due to the ghost of the $(1,2)$ 
orbit, and the period-$1$ oscillations at high magnetic field at
$\theta  = 11^\circ$ are related to the ghost of $(1,1)$ orbit.

Previously the ``ghost'' contibutions have been calculated using an extension
of the actual classical dynamics to the complex space. In this approach,
after a real periodic orbit ``dies'' in a bifurcation, a 
counterpart orbit continues to exist in complex space-time. The imaginary
part of the action of this complex periodic orbit describes the
exponential damping of it's contribution in the density of states
\cite{ghost_ref}.

Very recently, it has been shown, that complex dynamics can be
successfully used for the quantitative description of the tunneling
current\cite{Monteiro2}. A full stationary phase condition, 
when  applied directly to Eq.  (\ref{unclosed}), leads to {\it non-periodic}
and {\it complex} orbits. These trajectories were shown to smoothly
``interpolate'' between successive relevant periodic orbits, also 
providing an excellent quantitative description of the tunneling current in the ``ghost'' regime.

However, the approach of Ref. \cite{Monteiro2} does not possess one of the 
important advantages of the semiclassical theory described in the 
present work - the separation
 of the''isolated'' subsystems of the 
quantum well and  the emitter state.  In the approach of \cite{Monteiro2}
a different emitter state leads to a completely different set of complex 
trajectories,
which enter the final expression for the tunneling current.

The Closed Orbit Formula derived in the present paper ( Eq. (14) ) has the 
advantages that it only involves real orbits within the well, which depend on
the emitter state only weakly (through the location of the injection point
$y_i$).  It is therefore of interest to see if a semiclassical formula 
with this
rather different structure can also describe the ghost effect in the tunneling 
spectra quantitatively.    

The relations (\ref{closed}),(\ref{done_y}) provide a complete description
of all the ghost contributions in the tunneling spectra. For simplicity here
we will apply them to self-retracing orbits (the most relevant to experiment) 
and
introduce a further approximation to re-express the injected orbit in terms of
the closed orbit with the minimum $\Delta p_y$.

As it was argued in Section II.C, an injected orbit can substantially 
contribute 
to the tunneling current only if the change of momentum $\Delta p_y$ 
after returning
to the injection point is small.  Consider such an injected orbit when 
it is near
a tangent bifurcation as shown in Fig. \ref{fig:dpy_dy}.  
Just before the bifurcation there
exist no periodic orbit in a semiclassically 
small neighborhood of the injected orbit, 
but there does exist a 
unique nearby orbit of minimal 
$\Delta p_y$.  This orbit is 
determined by the classical dynamics 
in the well and not by the injection point, 
so we prefer to express the semiclassical 
tunneling current in terms of its properties, 
rather than those of the injected orbit.  
Let its point of contact with the emitter be 
at $y_0$, where $\left| y_0 - y_i\right|$ is 
semiclassically small by assumption; 
than the change of momentum
of the injected orbit\cite{self-retr} can be expressed as
\begin{eqnarray}
\Delta p_y & = & 
2 p_y^0 + K_0
\left( y_i - y_0 \right)^2,
\ \ \ \ 
K_0 \equiv \left. \frac{\partial^2 p_y}{\partial y^2} \right|_{y = y_0}
\label{s_ghost}
\end{eqnarray}
Assuming that the electrons in the emitter (or collector) are in the lowest Landau level, 
for the corresponding ghost contribution to the tunneling rate we obtain:
\begin{eqnarray}
W_{\rm ghost} = A^{\rm sc}_{\rm gh} 
\exp\left(- \frac{t_0}{\tau_{\rm opt}} \right)
\cos\left( \frac{S_0}{\hbar} + \phi_0\right)
\label{w_ghost}
\end{eqnarray}
where
\begin{eqnarray}
A^{\rm sc}_{\rm gh} & = & A_0 \exp\left( - \xi^2\right)
\label{eq:Agh} 
\end{eqnarray}
\begin{eqnarray}
A_0 & = & 
\frac{4}{m^*} 
\frac
{p_z^0}
{
\left[
1 + 
\left(\frac{m^0_{12} \hbar}{ 2 l_B^2} \right)^2
\right]^{1/4} }
\exp\left[ - \left(p_y^0 l_B/\hbar\right)^2 \right],
\label{eq:A0}  
\end{eqnarray}
\begin{eqnarray}
\xi & = & 
\left[
\frac
{
\left(
p_y^0 + K_0
\left( y_i - y_0 \right)^2
\right)^2
}
{
1 + 
\left(
K_0
\left(y_i - y_0\right) \frac{l_B^2}{2 \hbar}
\right)^2
}
-
\left(p_y^0\right)^2 
\right]^{1/2}
\frac{l_B}{\hbar},
\label{eq:F0}
\end{eqnarray}
$m_{ij}^0$, $t_0$ and $S_0$ are respectively the monodromy matrix, 
the traversal time and the action of the
closed orbit with minimum change of momentum $\Delta p_y$, and
the $\phi_0$ is the corresponding phase, which, in addition to the 
constant part, related to  the topological
index of the orbit, has also a smoothly varying component. Although
it is 
straightforward to obtain the latter from (\ref{done_y}), the result
is somewhat complicated, and is not presented here. 
Note, that the classical input for the Eq. (\ref{w_ghost}) depends again only on the
injection point and not on 
the particular emitter state.

Thus in  Eq. (\ref{w_ghost}) we have derived a relatively simple explicit formula 
for the ghost contributions in the tunneling spectrum
in terms of the classical properties of the closed orbit with minimal values 
of $\Delta p$ ($\equiv 2 p_y$ for a self-retracing closed orbit).  Note that 
the strength of the ghost contribution is exponentially small in 
$\left( \Delta p / p_B\right)^2$,
where $p_B = \hbar /l_B \sim \sqrt{\hbar}$.

Now we proceed with a detailed numerical test of the ``ghost formula''
(\ref{w_ghost}).
Note that in principle the expression (\ref{w_ghost}) can describe
the tunneling rate both 
from the ground Landau level in the emitter to the well  
{\it or} from the well to the ground Landau level in the collector. 
Although the actual experimental tunneling
current in the incoherent regime does not depend on the latter quantity, 
either can be used
to test the validity of Eq. (\ref{w_ghost}) in the ghost regime.
In fact the tunneling rate into the collector is a more convenient 
object for numerical 
computation since it is non-zero for energies less than $eV$ and thus can be 
expressed in terms of lower-lying eigenstates for interesting value of $eV$. 
 Effectively
then we will test the formula in the single-barrier model which we 
studied in detail in 
reference \cite{tiltprb} for which the classical dynamics 
depends only on $\beta$.
The first (as a function of $\beta$) relevant bifurcation 
in that model is the tangent bifurcation of the period-one 
orbits which we designated $(0,1)^{+(1)}$ and $(0,1)^{-(1)}$ 
(for a detailed description of the corresponding
 classical dynamics, see Ref. \cite{tiltprb}, Section III.C).
We will now consider the ``ghost'' related to that tangent bifurcation 
at $\theta = 11^\circ$.  Although for any given system parameters $y_i$ and $y_0$ are fixed in Eq. (\ref{w_ghost}), for a complete test of the formula we treat
$y_i$ as an independent variable and study the amplitude $A^{\rm sc}_{\rm gh}$
as a function of both $y_i$ and $\beta$ ($\sim \sqrt{\varepsilon}$).

In Fig. \ref{fig:a0} we show the comparison of the exact and
 semiclassical results for $A_0$.  With {\it no fitting parameters} the 
semiclassical calculation (represented by a dashed line
in Fig. \ref{fig:a0}) is in a good agreement
with exact data (open circles). However, essentially perfect agreement can
be obtained when one introduces a single fitting parameter by means of
the definition
$A^{\rm sc}_0\left(\varepsilon_L\right)$
where the ``longitudinal energy'' $\varepsilon_L$ is defined as 
$\varepsilon_L \equiv \varepsilon - \kappa \hbar \omega_c$,
with the $\kappa \lesssim 1$ being the fitting parameter.
 
The quantity $\kappa \hbar \omega_c$ can be interpreted as the energy 
of the motion transverse to the direction of the periodic orbit.
As was 
shown in Section II.C, a careful consideration of the higher-period orbits, 
related to the repetitions of the short trajectories, leads to the
replacement of the 
total energy by it's longitudinal part in the {\it phase} of the 
corresponding contribution
to the tunneling rate. It is natural to assume, that 
such a replacement should 
take place in {\it both} the phase {\it and} the amplitude. 
However, because the transverse 
energy $\varepsilon_\perp \sim \hbar$, the correction to the amplitude 
is of the next
order in $\hbar$, and is not captured by the semiclassical theory. 
The calculation, presented
in Fig. \ref{fig:a0} gives a clear indication that the amplitude 
should also, in fact, depend
on the {\it longitudinal} energy.

 As follows from (\ref{w_ghost}), neglecting the 
slowly varying prefactor, the amplitude depends only on the ``scaled
deviation'' from it's maximum value $\xi$:
\begin{eqnarray}
A^{\rm sc}_{\rm gh}\left(y_i, \beta\right) 
\sim \exp\left[-\xi\left(y_i,\beta\right)^2
\right]
\label{eq:axi}
\end{eqnarray}
In Fig. \ref{fig:ay} we plot the scaled amplitude as a function of $\xi$ for 
eight different values of $\beta$, and a good agreement is found with the
predicted form (\ref{eq:axi}). Note, that since the formula ({\ref{w_ghost}) 
is,
in turn, also based on the expansion near the contact point $y_0$ of the 
``minimum $\Delta p_y$ closed orbit'', the deviation of the exact data from 
(\ref{eq:axi}) should increase for larger $\xi$, again in agreement
with the calculation presented in Fig. \ref{fig:ay}.

\section{Conclusions}

In the work which we have concluded with the present paper,
we have developed a complete semiclassical theory
of resonant tunneling in quantum wells in a tilted magnetic field. The 
theory was shown to be in good qualitative and quantitative agreement
with the experiments of Muller et al \cite{Bell}, and explains many puzzling 
features of the 
tunneling spectra.  In particular we have explained the phenomenon of persistent
scarring of wavefunctions in terms of the properties of cusp bifurcations and 
bifurcations cascades, which lead to metastable periodic orbits.  We expect
these concepts to extend to other systems and for the same persistent 
scarring to be found in such systems in the appropriate parameter ranges.
Furthermore we have demonstrated explicitly the possibility of describing 
quantitatively the
ghost contributions of vanished periodic orbits in terms of a semiclassical 
theory of real closed orbits.

The authors wish to thank G.~Boebinger, D.~Shepelyansky, T.~S.~Monteiro, 
T.~M.~Fromhold, H.~Mathur, G.~Hackenbroich and J.~Delos for helpful
discussions. We further thank T.~S.~Monteiro for communicating the
results of recent work on the semiclassical formulas and their extension
to complex orbits prior to publication.  
This work was partially supported by NSF Grant No. DMR-9215065.

\appendix

\section{Semiclassical Green function near a rectangular barrier}

The equation (\ref{w_green}), used for the calcluation of the 
tunneling rate from the emitter to the well, requires the 
knowledge of the normal derivative of the Green function
$\partial^2_{z_1 z_2 } G\left(y_1, z_1 = 0; y_2, z_2 = 0\right)$.  
Because of the apparent discontinuity of the 
potential {\it exactly} where the Green function 
is evaluated, the semiclassical approximation for this object
should be performed with care.

First, we consider the first derivative $\partial_{z_1} 
G(y_1, z_1=0; y_2, z_2)$, which we define as
\begin{eqnarray}
\frac{\partial G\left(y_1, z_1 = 0; y_2, z_2\right)}
{\partial z_1 }
 & = & 
\lim_{z_1 \to 0}  
\frac{
\partial G\left(y_1, z_1; y_2, z_2\right) 
-
G\left(y_1, 0; y_2, z_2\right)}
{z_1} 
\end{eqnarray}
The value $z_1 \to +0$ should be {\it classically small} 
(e.g. $z_1 \sim \hbar$), so that the change 
of the smooth part of the total potential $V(y,z)$
at the distance $z_1$
is small compared to the kinetic energy of the electron.

When $z_1 \ne 0$, the semiclassical Green function can be expressed as
a sum over classical trajectories \cite{Gutzwiller} (indexed by $\nu$) 
connecting the points $(y_1,z_1)$ and $(y_2,z_2)$:
\begin{eqnarray}
G & = & 
\frac{2 \pi}{\left(2 \pi i \hbar \right)^{3/2}}
\sum_\nu
\sqrt{\left|D_\nu\right|} 
\exp\left( \frac{i}{\hbar} S_\nu - \frac{i \pi \mu_\nu}{2} \right)
\label{gsc}
\end{eqnarray}
where $D$ is the (semi)classical amplitude, $S$ is the action integral and
$\mu$ is the topological index of the trajectory $\nu$.

When the  starting point $(y_1,z_1)$ is near the emitter barrier, the 
Green function can be expressed in terms of the classical 
trajectories starting {\it at} the emitter barrier - i.e. from the point 
$(y_1,0)$. For each such trajectory $j$ there are {\it two} trajectories
$j_1$ and $j_2$ [starting from $(y_1,1)$] with different number of 
collisions with the barrier, 
which collapse to $j$ when $z_1 \to 0$ - see Fig. \ref{fig:app}. In the 
semiclassical limit we have:
\begin{eqnarray}
S_{j_1} & = & S_{j} - p_z^{(1)} z_1, \ \ \ 
S_{j_2}  =  S_{j} + p_z^{(1)} z_1, 
\nonumber \\
\mu_{j_1} & = & \mu_j, \ \ \ 
\mu_{j_2}  =  \mu_j + 2 
\label{j12j}
\end{eqnarray}
where ${\bf p}^{(1)}$ the electron momentum at
the starting point, and 
the extra $2$  in the expression for $\mu_{j_2}$ corresponds
the phase increment of $\pi$ 
due to an additional (with respect to the trajectory $j$) reflection
from the hard wall.

Substituting (\ref{j12j}) into (\ref{gsc}), 
we obtain:
\begin{eqnarray}
G\left(y_1,z_1;y_2,z_2\right) 
& =  & 
\frac{4 \pi i}{\left(2 \pi i \hbar\right)^{3/2} }
\sum_j
\sqrt{\left|D_j\right|} 
\sin\left(\frac{p_z^{(1)} z_1}{\hbar}\right)
\exp\left( \frac{i}{\hbar} \ S_j - \frac{i \pi \mu_j}{2} \right)
\label{gsc1}
\end{eqnarray}
and therefore the first normal derivative 
\begin{eqnarray}
\left.
\frac{\partial G\left(y_1,z_1;y_2,z_2\right)}
{\partial z_1}\right|_{z_1 = 0}  
& = &  
\frac{4 \pi i}{\left(2 \pi i \right)^{3/2} \hbar^{5/2} }
\sum_j p_z^{(1)} 
\sqrt{\left|D_j\right|} 
\exp\left( \frac{i}{\hbar} \ S_j - \frac{i \pi \mu_j}{2} \right)
\label{gscd1}
\end{eqnarray}

Using a similar approach, one can show, that 
the second derivative
\begin{eqnarray}
\left.
\frac{\partial^2 G\left(y_1,z_1;y_2,z_2\right)}
{\partial z_1 \partial z_2}\right|_{z_1 = z_2 = 0} 
 =  
\frac{8 \pi i}{\left(2 \pi i \right)^{3/2} \hbar^{7/2} }
\sum_j p_z^{(1)} p_z^{(2)} 
\sqrt{\left|D_j\right|} 
\exp\left( \frac{i}{\hbar} \ S_j - \frac{i \pi \mu_j}{2} \right)
\label{gscd2}
\end{eqnarray}
where   ${\bf p}_z^{(1)}$ and  ${\bf p}_z^{(2)}$ are respectively
the initial
and the final momenta.

\begin{figure}[htbp]
\begin{center}
\leavevmode
\epsfbox{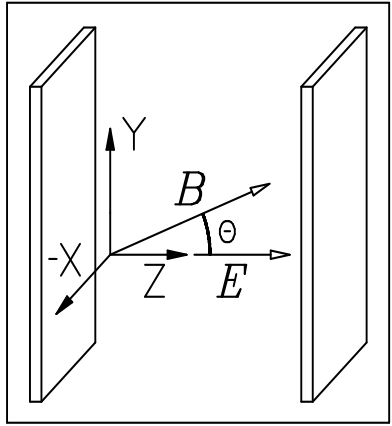}
\end{center}
\caption{Schematic of the geometry and coordinate system used in this work}
\label{fig:schematic}
\end{figure}

\begin{figure}[htbp]
\begin{center}
\leavevmode
\epsfbox{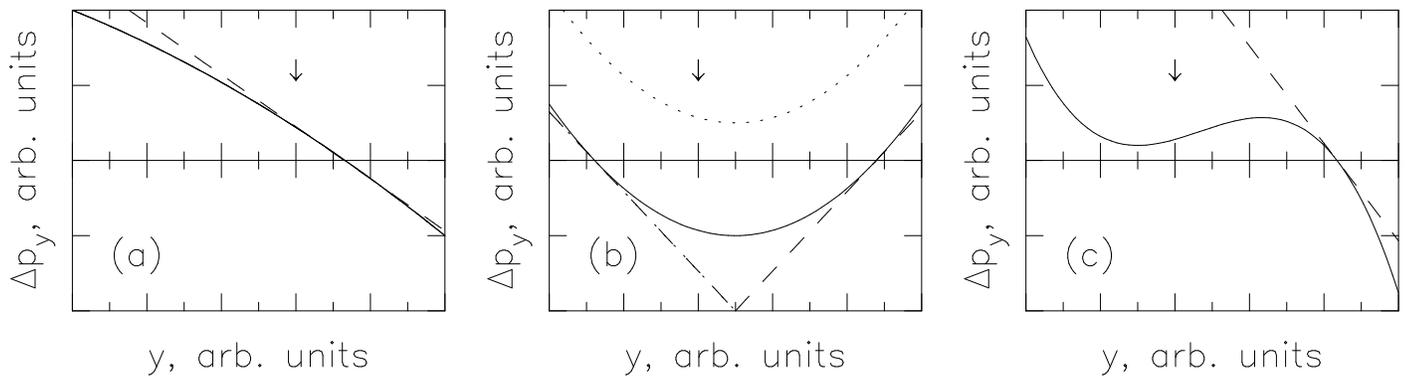}
\end{center}
\caption{The change of momentum $\Delta p_y$ for the closed
orbits of a particular family,
as a function of ``starting coordinate'' $y$. The dashed lines show
the linear approximation for $\Delta p_y$, used in 
(\protect\ref{eq:S:clo-po}),
(\protect\ref{eq:p:clo-po}). The arrow indicates the 
``injection point'' $y = y_i$. If at $y = y_i$ the value of 
$\Delta p_y$ is small, the linear approximation is justified
(a), but fails in the vicinity of a bifurcation (b),(c). In (b) the 
dotted and solid lines respectively represent the 
behavior before and after a tangent bifurcation.}   
\label{fig:dpy_dy}
\end{figure}

\begin{figure}
\begin{center}
\leavevmode
\epsfbox{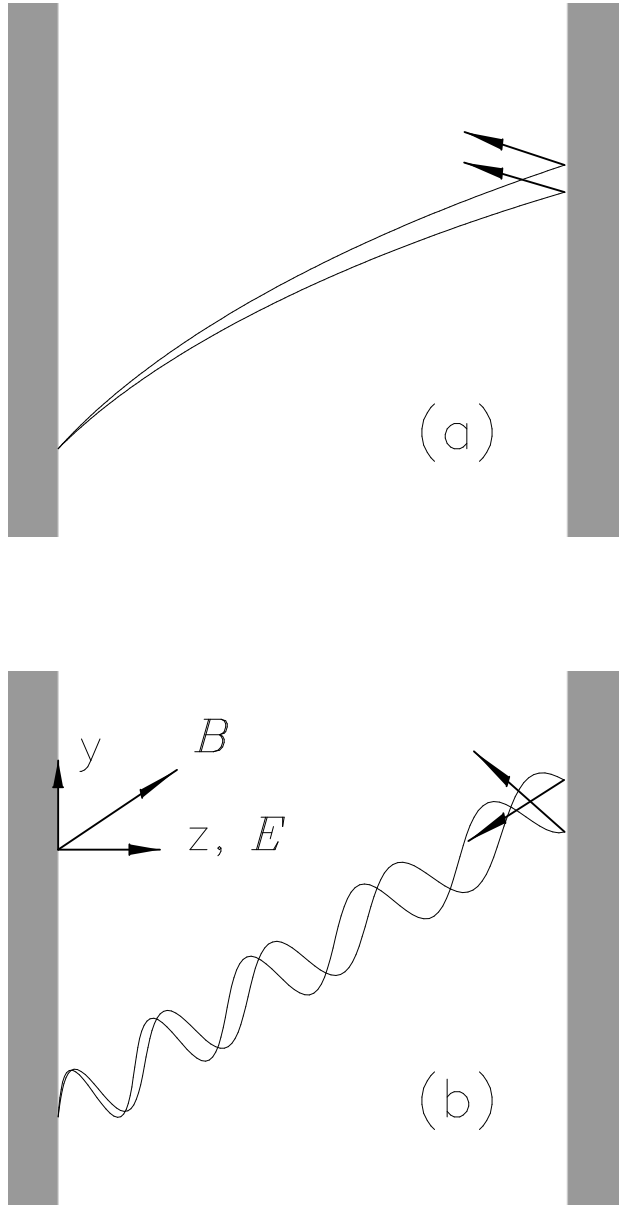}
\end{center}
\caption{
The evolution of two classical trajectories, started
from the same point at the emitter barrier with 
almost identical velocities, for (a) 
$\omega_c \tau \protect\lesssim 1$,
and (b) 
$\omega_c \tau \gg 1$, where 
$\tau$ is the well traversal time. The black arrows 
represent the velocities immediately after collision
with the collector barrier; the inset shows the 
directions
of the magnetic and electric fields, and the
coordinate system. Note the large amplification of small differences
when the cyclotron rotation rate is large compared to the traversal time.  
This is the origin of chaos in the system, as discussed in the text.
}
\protect\label{fig:origin_of_chaos}
\end{figure}

\begin{figure}
\begin{center}
\leavevmode
\epsfbox{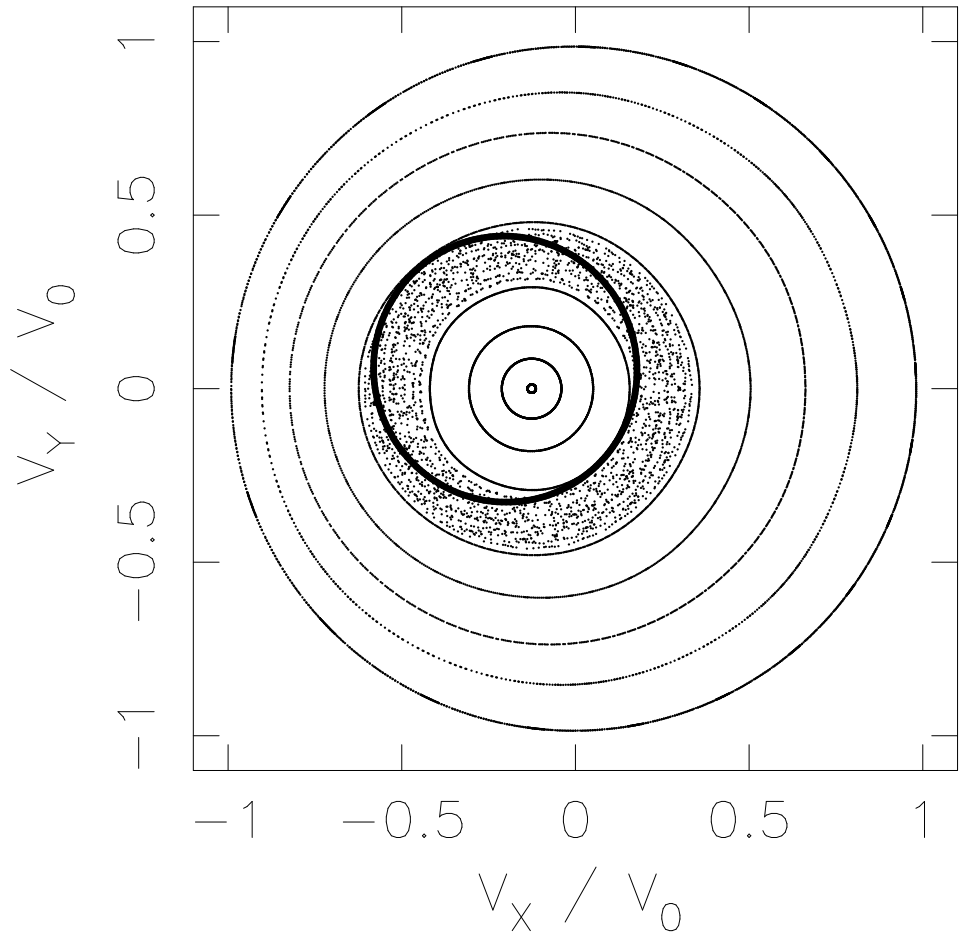}
\end{center}
\caption{The ``chaotic halo'' near the critical boundary (thick solid line),
created by nonanaliticity of the Poincare map, seen at the collector 
barrier surface of section at $\beta = 2$, $\gamma = 1.17$ and $\theta =
30^\circ$.  Note that the halo region in which trajectories can hit the 
emitter barrier intermittently is highly chaotic while the inner and 
outer regions, where trajectories either always hit or always miss the 
emitter, are still quite regular.  All the periodic orbits relevant
to the tunneling current are born at the critical boundary and then
 move inwards.}
\label{fig:halo}
\end{figure}

\begin{figure}
\begin{center}
\leavevmode
\epsfbox{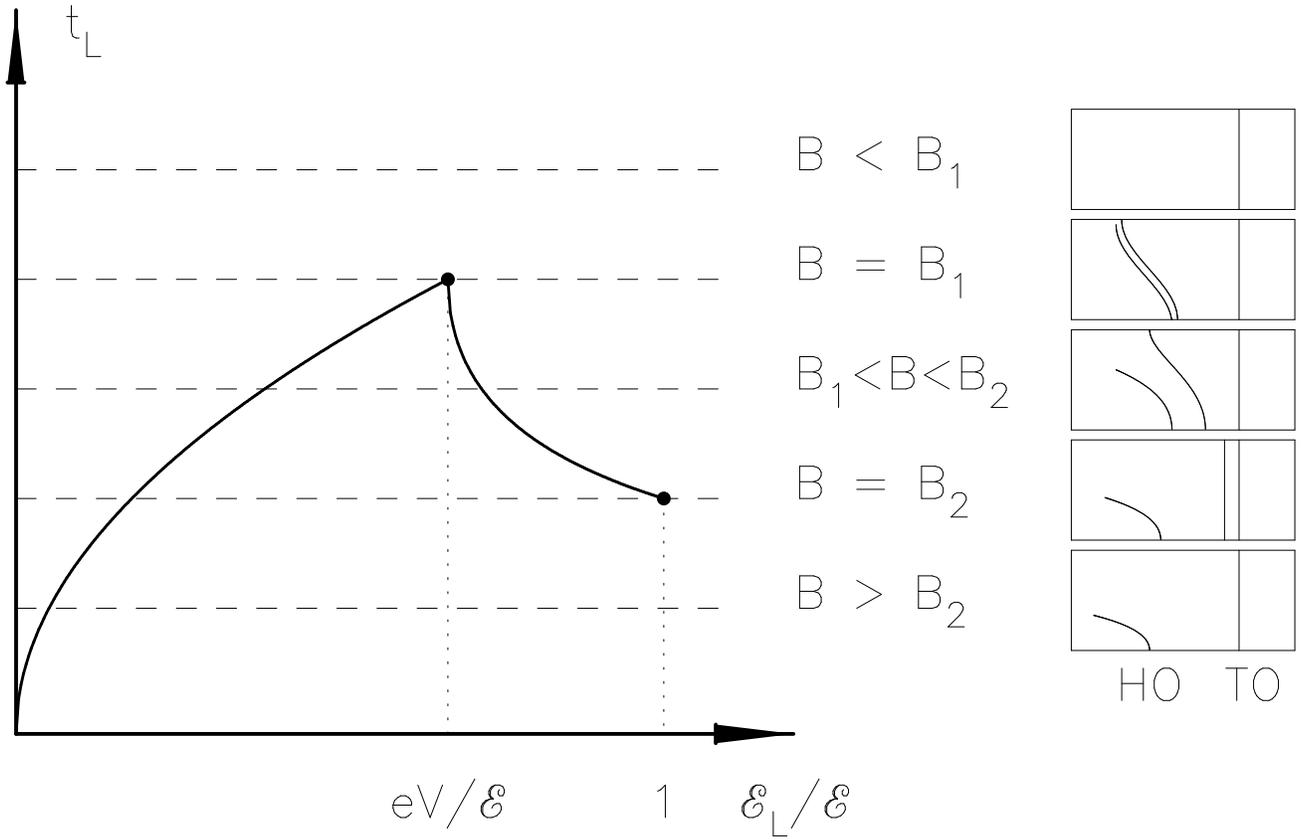}
\end{center}
\caption{
The dependence of the traversal time of longitudinal motion
$t_L$ on the fraction of the total energy which is longitudinal,
$\varepsilon_L/\varepsilon$. The dashed lines
represent the three different regimes discussed in the text, $B<B_1$,
$B_1<B<B_2$, $B>B_2$ and the corresponding borders
$B = B_1$ and $B=B_2$. The insets on the right
schematically represent the different periodic
orbits in each regime. 
\label{fig:tl_e}
}
\end{figure}

\begin{figure}
\begin{center}
\leavevmode
\epsfbox{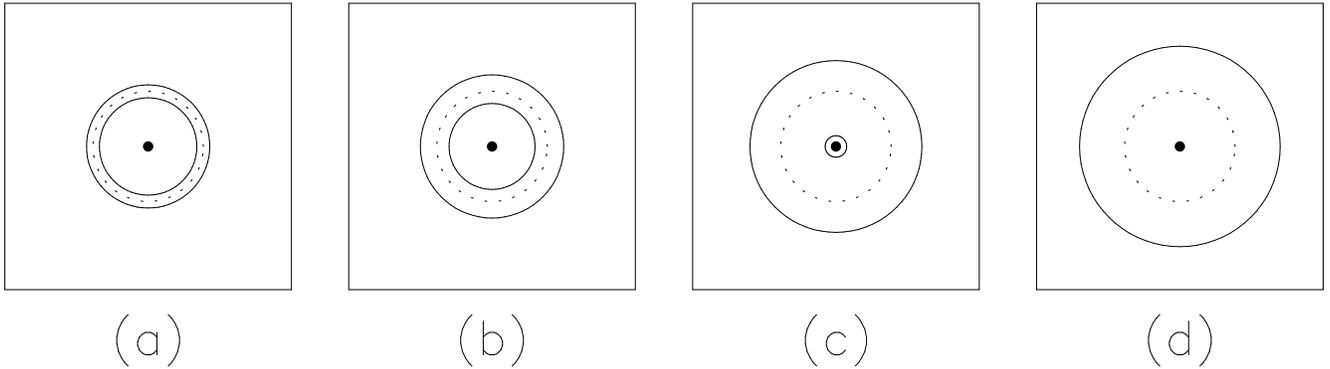}
\end{center}
\caption{
The schematic representation of the tori at the Surface
of Section at 
(a) $B = B_1 + \epsilon$, 
(b) $B_1 < B < B_2$, 
(c) $B = B_2 - \epsilon$,
(d) $B > B_2$, $\epsilon \to +0$.
Dotted circle represents the critical boundary at $\theta=0$.
\label{fig:schem_psos}
}
\end{figure}

\begin{figure}
\begin{center}
\leavevmode
\epsfbox{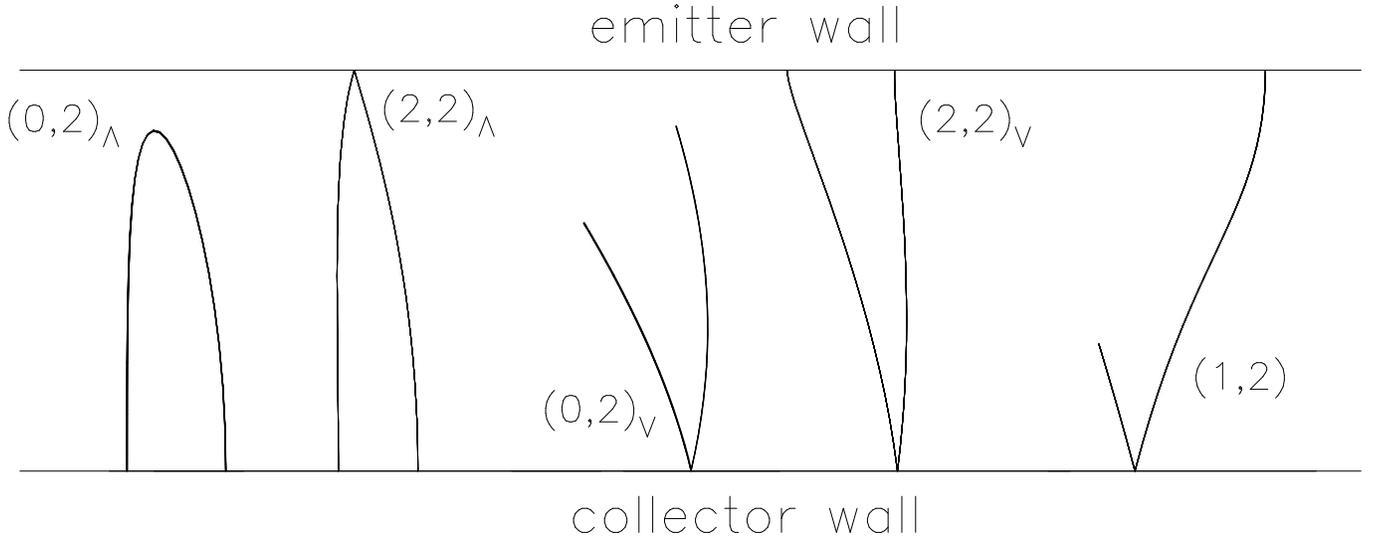}
\end{center}
\caption{
Examples of the different types of period-$2$ orbits in $y$-$z$ projection.
At higher magnetic fields similar orbits arise with one or more extra
turns around the magnetic field direction.
\label{fig:po2}
}
\end{figure}

\begin{figure}
\begin{center}
\leavevmode
\epsfbox{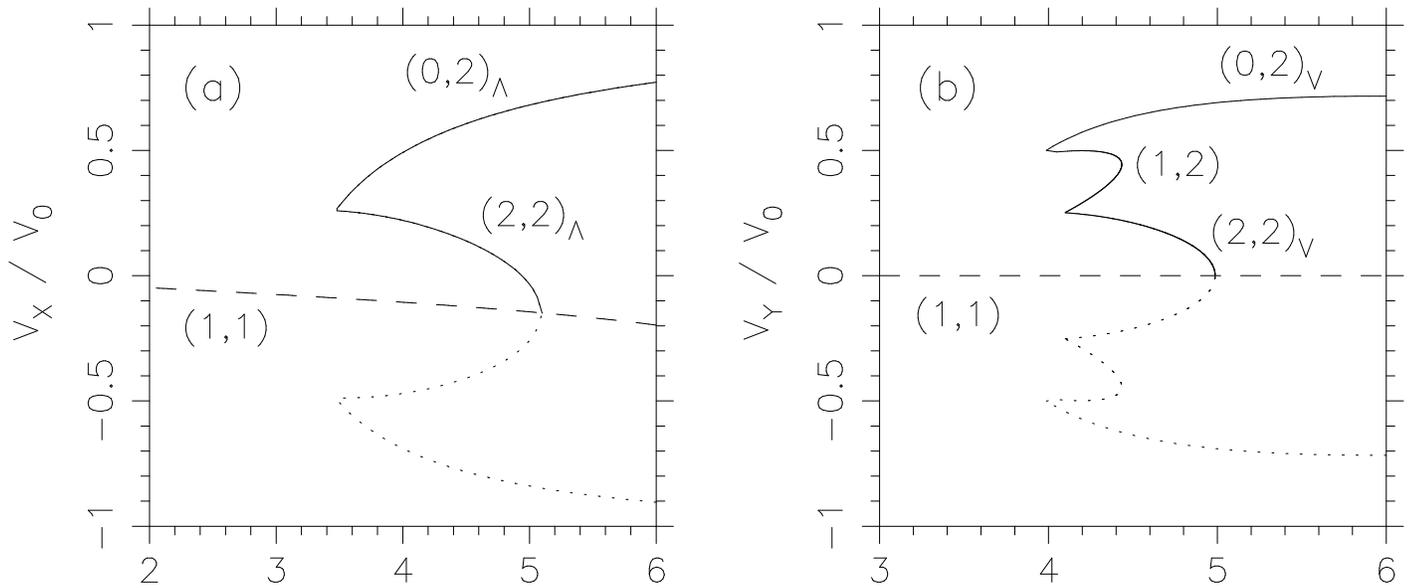}
\end{center}
\caption{
The bifurcation diagrams of the period-$2$ $\Lambda$-orbits
(a) [$\theta = 11^\circ$] and $V$-orbits (b) 
[$\theta = 17^\circ$]. Each of the period-$2$ orbits is 
represented by it's velocity [$v_x$ for $\Lambda$-orbits and 
$v_y$ for $V$-orbits] at the 
collisions with the collector barrier,
with the velocity at the first collision shown by solid line, and
the velocity at the second bounce shown by dotted line.
The dashed line represents 
the period-$1$ orbit $(1,1)$. 
\label{fig:bif_cascade}
}
\end{figure}

\begin{figure}
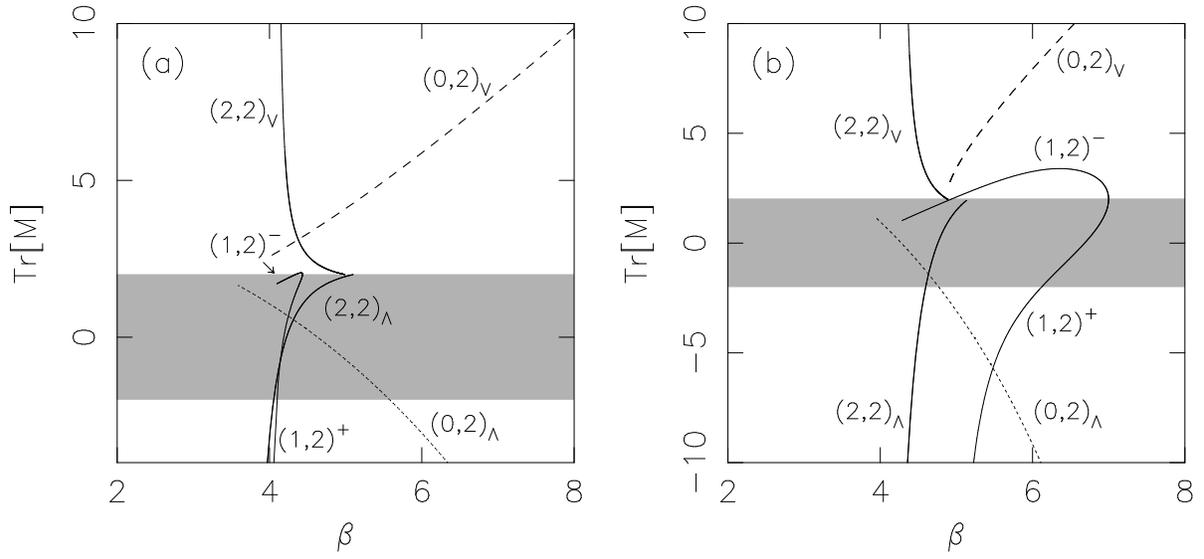

\begin{center}
\leavevmode
\epsfxsize = 3in
\epsfbox{fig08a.epsi}
\hskip 0.5 truecm
\leavevmode
\epsfxsize = 3in
\epsfbox{fig08b.epsi}
\end{center}
\caption{
The trace of the monodromy matrix 
for different period-$2$ orbits at
(a) $\theta = 17^\circ$ and (b)
$\theta = 28^\circ$. 
{\it All} these period-two orbits
appear in cusp bifurcation.
One
cusp bifurcation produces the 
$\Lambda$-orbits $(0,2)_\Lambda$ and $(2,2)_\Lambda$,
another cusp bifurcation generates the
orbits $(1,2)^-$ and $(2,2)_V$,
and yet another cusp bifurcation produces
$(0,2)_V$ and $(1,2)^+$. Note the 
discontinuity of the trace of the monodromy
matrix of the ``more connected'' partners
$(2,2)_\Lambda$, $(2,2)_V$ and $(1,2)^+$.
The single ``metastable'' orbit $(1,2)^-$
leads to a strong scarring, it's interval
of existence substantially increases with the
tilt angle.
\label{fig:trM}
}
\end{figure}

\begin{figure}
\begin{center}
\leavevmode
\epsfbox{fig09.epsi}
\end{center}
\caption{
A wavefunctions scarred by unstable
$(1,2)^-$ orbit with $y$-$z$ projections of orbit superimposed (a),
and the ``Scar strength'' $H$ vs. the 
scaled action $S(\varepsilon_n)/h$ of the $(1,2)^-$ orbit
which scars the eigenstate of energy $\varepsilon_n$.  
The arrow indicates the value of $\beta$ for the tangent bifurcation, 
which give birth to the periodic orbit. Note that when increasing
action (or energy) the tangent bifurcation at higher $\protect\beta 
\protect\sim 1/\protect\sqrt{\protect\varepsilon}$ 
occurs at the {\it lower} action side. The peaks of the scar strength 
below the tangent bifurcation are due to the ``ghost effect'' 
\protect\cite{ghost_ref,ghost}. 
Scaled action below the bifurcation points
is obtained by linear extrapolation of the 
(approximately linear) function $S(\varepsilon)/h$.  
``Scar strength'' $H$ is defined as the 
value of the Husimi function
$H(y^0,p_y^0) \equiv 
\int dy \ dp_y W_n(y,p_y) 
\exp\left(-(y-y^0)/l_B^2 - l_B^2 (p_y - p_y^0)^2/\hbar \right)
$ 
calculated for the normal derivative of the wavefunction 
( $W_n(y,p_y) = h^{-1} \int d\Delta y \ 
\partial_z \Psi\left(y - \Delta y/2,0\right)
\partial_z \Psi\left(y + \Delta y/2,0\right)
\exp\left( i p_y \Delta y / \hbar \right)
$
)
at the location of the fixed point $(y^0 \propto v_x, p_y^0 \propto  v_y)$
of the unstable periodic orbit.
\protect\label{fig:scar_strength}
}
\end{figure}

\begin{figure}
\begin{center}
\leavevmode
\epsfbox{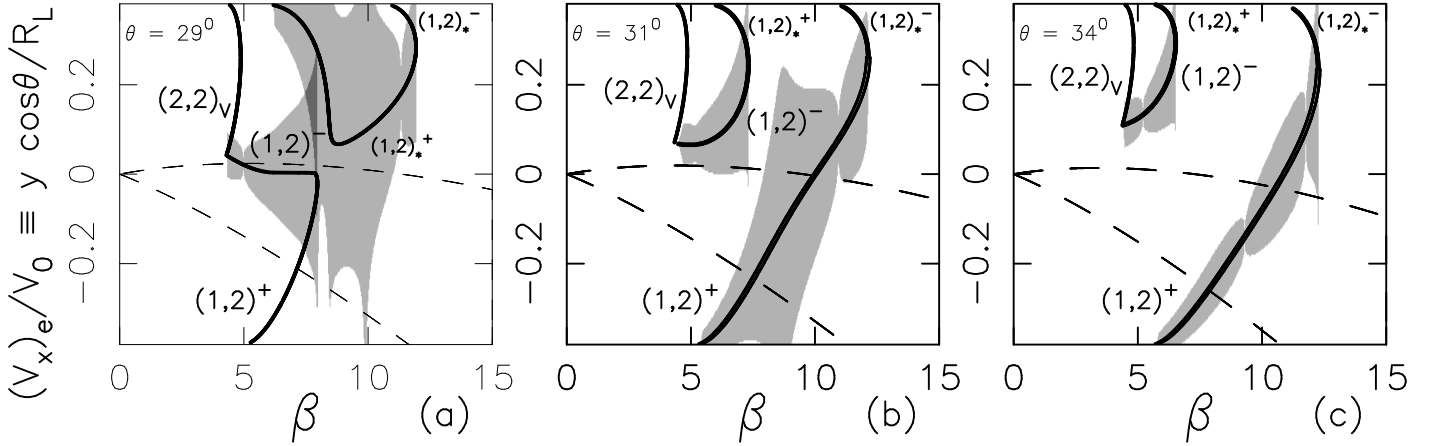}
\end{center}
\caption{
Bifurcation diagrams for the four period-two orbits relevant to the
peak-doubling in the interval $\theta=29^{\circ}-34^{\circ}$ (see text).
Thick lines indicate y-coordinate of the single collision with the
emitter; these lines coincide at bifurcations.  $\theta = 29^{\circ}$
(a) indicates behavior before the exchange bifurcation;
($\theta = 31^\circ$ (b) and $34^\circ$ (c) after.
Shading represents the localization lengths $l_{\mu}$ associated
with the relevant orbits [$(1,2)^-$ and $(1,2)_*^+$ for (a), and
$(1,2)^-$ and $(1,2)^+$ for (b),(c)].  Region between the dashed lines
denotes
semiclassical width of emitter state; overlap indicates a large
contribution to the tunneling current.
\label{fig:exchange_cascade}
}
\end{figure}

\begin{figure}
\begin{center}
\leavevmode
\epsfbox{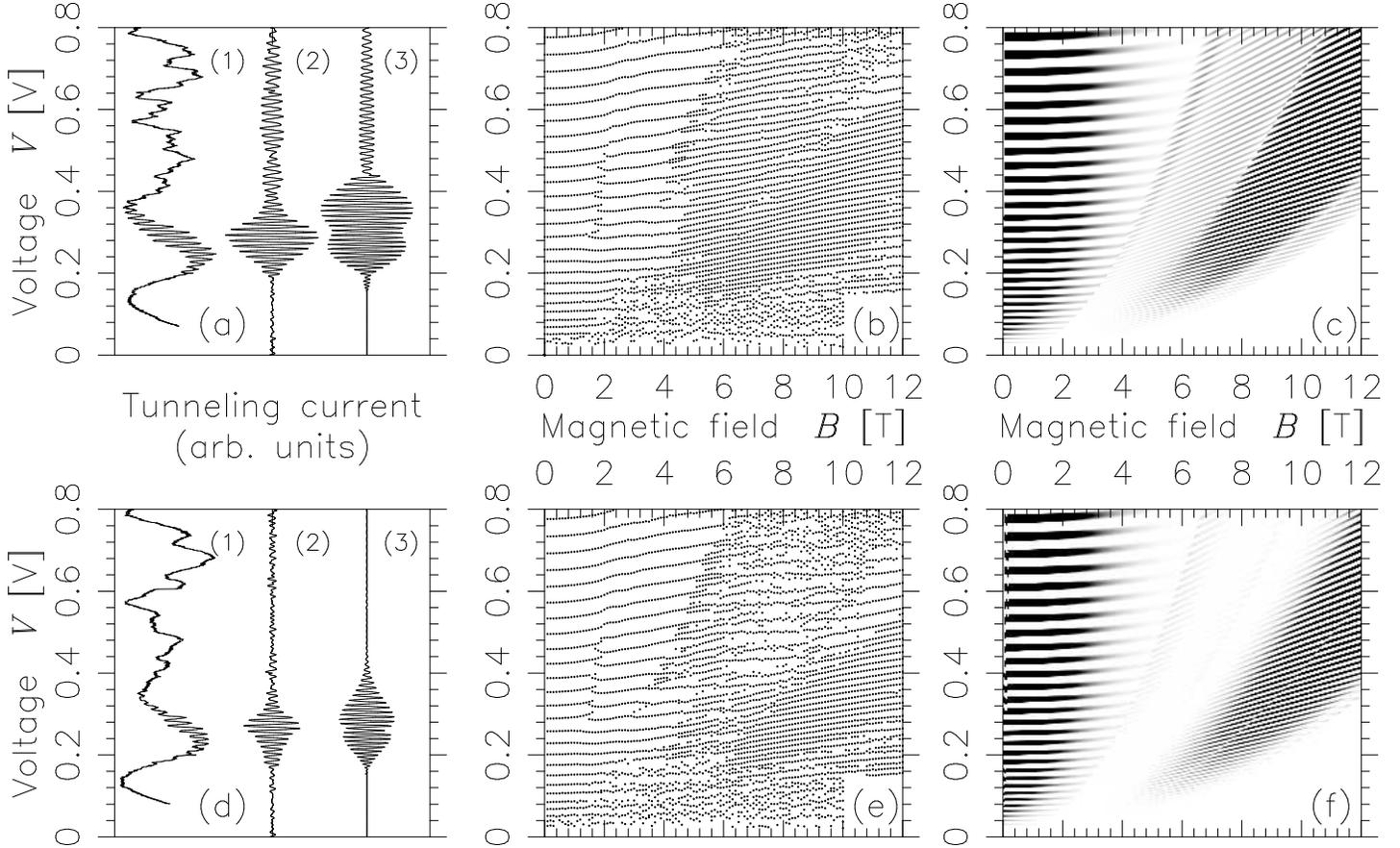}
\end{center}
\caption{
Figs. (a),(d): resonant tunneling $I-V$ traces for $\theta = 31^{\circ},
\theta = 34^{\circ}$ at $B=8$T.  Trace (1) is raw experimental data, trace (2)
is same data, filtered to retain only period-two oscillations, trace (3) is
semiclassical prediction from 
Eq. (\protect\ref{w_po}). The modest discrepancies 
in the shape of the
envelope of the amplitude of the oscillations are due to the
inaccuracy of the quadratic semiclassical theory near the
bifurcation which occurs around 0.3V at 8T. 
Figs. (b),(e): peak positions vs. voltage and magnetic field,
determined from multiple sets of experimental $I-V$ data at
$\theta = 31^{\circ},34^{\circ}$.
Figs. (c),(f): semiclassical I-V oscillations for same, note gray-scale
indicates relative amplitudes, not just peak positions.
Note disappearance of high voltage oscillations at $\theta = 34^{\circ}$
due to movement of $(1,2)^-$ orbit away from accessibility after the
exchange bifurcation (see Figs. 
\protect\ref{fig:exchange_cascade}b,
\protect\ref{fig:exchange_cascade}c 
and text) [The Figure is reprinted from Ref.\protect\cite{tiltsc}].
\label{fig:BVplots}
}
\end{figure}

\begin{figure}
\begin{center}
\leavevmode
\epsfysize = 2.5 in
\epsfbox{fig1clf.epsi}
\end{center}
\caption{
The dependence of the amplitude of 
``ghost'' oscillations on $\beta$
for $\theta = 11^\circ$. The open circles
represent the results of the exact calculation.
The dashed line corresponds to the 
semiclassical formula (\protect\ref{eq:A0}).
The solid line corresponds to the semiclassical 
calculation, which takes into account the
transverse energy $\kappa \hbar \omega_c$, 
with $\kappa = 0.55$.
This is in a good agreement with 
a simple estimate of $\kappa \approx 0.42$ 
from the transverse energy of the
{\it stable} $(0,1)^+$ orbit, which is soon
to be born in the tangent bifurcation.
}
\label{fig:a0}
\end{figure}

\begin{figure}
\begin{center}
\leavevmode
\epsfysize = 2.5 in 
\epsfbox{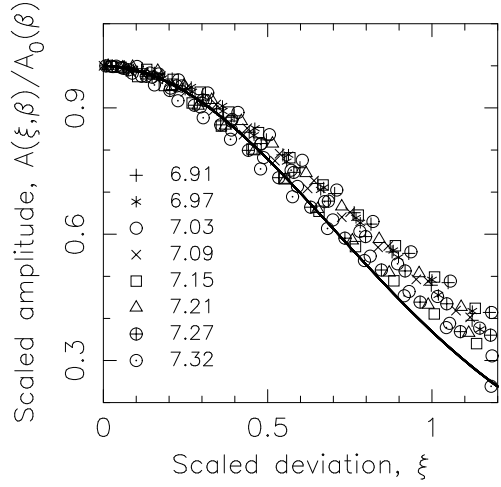}
\end{center}
\caption{
The dependence of the scaled amplitude 
on the scaled deviation $\xi$ from it's
maximum value. The solid line represents
the functional form (\protect\ref{eq:axi}).
Different simbols represent the results of the 
exact calculation for different $\beta$'s, with the
corresponding values shown in the inset. Since
the result (\protect\ref{eq:axi}) is based on the 
expansion near $\xi = 0$, the deviations from the
semiclassical result increase for larger $\xi$.
}
\label{fig:ay}
\end{figure}

\begin{figure}
\begin{center}
\leavevmode
\epsfysize = 2.5 in 
\epsfbox{fig_app.epsi}
\end{center}
\caption{
The classical trajectories, which
contribute to $G\left(y_1, z_1; y_2,z_2\right)$
for $z_1 \to +0$.
}
\label{fig:app}
\end{figure}

\begin{references}
\bibitem{narrow} Although the ``width'' of such regions in the 
parameter space is ``semiclassically'' small, it was shown
\protect\cite{Monteiro}, that 
at $\theta = 27^\circ$ the ``ghost'' effect is responsible for a
substantial part of the experimentally seen peak-doubling.
\bibitem{explain_where_U0_is} In this representation, the dependence of the 
matrix element on the barrier height $U_0$ is carried by the wavefunctions
of the isolated emitter, calculated at $z = 0$.
\bibitem{BR0} This stationary phase integration 
was however used by Bogomolny and Rouben in 
\protect\cite{BR} to obtain a simplified analytical formula
for the tunneling current. In Ref. \cite{Saraga} this formula
was shown to give only poor agreement with exact calculations
and experimental data as we would expect from the arguments presented
in the text.
\bibitem{bifurcations}  The semiclassical formula of Eq. (\ref{w_po}) 
breaks down at bifurcations, when the periodic orbits are not isolated.
Correspondingly, the semiclassical width can either diverge or
vanish (depending on the nature of the bifurcation).  These divergences
are spurious but are integrable in Eq. (\ref{w_po})
and only lead to inaccuracies in a small
interval of $\beta$ very close to the bifurcation. In these intervals
one should use the Closed Orbit Formula.
\bibitem{scalings} This representation was introduced in Ref.
\protect\cite{tiltprb}.  Alternative scaling relations were used in
Refs. \protect\cite{Bell,ss,Monteiro}.
\bibitem{period-three} 
Unlike their birth in cusp bifurcations, all emitter orbits ``die'' in 
analytic backwards bifurcations which follow the generic rules of 
hamiltonian 
bifurcation theory.  Hence, for example, backwards bifurcations with the 
period-one traversing orbit is allowed for period-two
orbits, but not for period-three orbits, which vanish near the traversing 
orbit in a tangent bifurcation according to the ``touch-and-go'' scenario 
described in ref. \cite{Delos}.
\bibitem{dss} The definition of the ``scar strength'' is not unique, e.g.
one may instead integrate the modulus of the wavefunction along the
periodic orbit in the real space\protect\cite{Wintgen}. The difference in the 
definition, however, will not affect the qualitative difference between 
``localized'' and ``ergodic'' states.
\bibitem{explain_why_py_irrelevant} Note, that for all these orbits 
at the point of collision with the emitter barrier $p_y = 0$.
\bibitem{endnote}
The major remaining discrepancy, that the experimental
peak-doubling regions are observed at lower voltages than calculated
by the semiclassical theory, arises from an offset voltage in the
sample\protect\cite{Bell} and band non-parabolicity effects
\protect\cite{Fromhold_prbrc}.
\end{references}
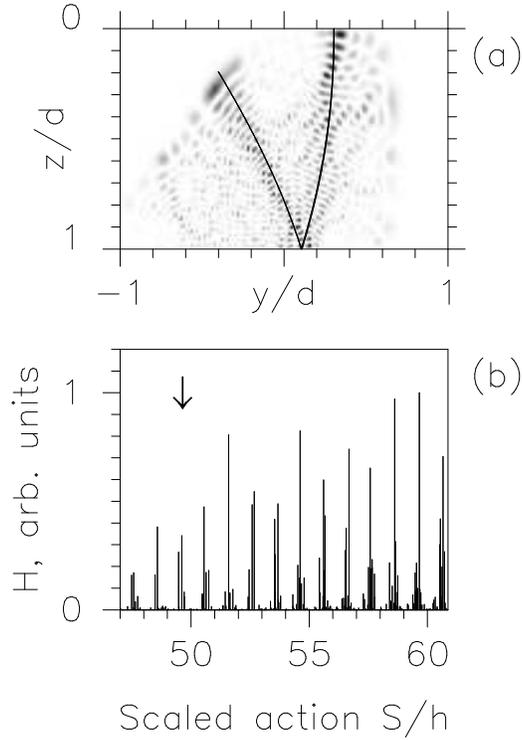
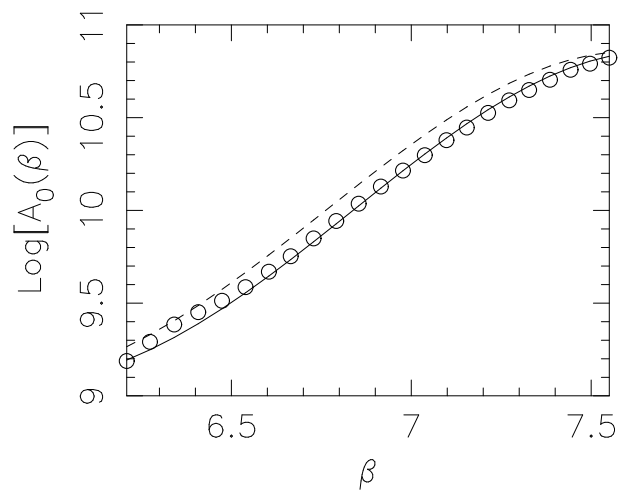
\end{document}